\documentclass[preprint,showpacs,preprintnumbers,amsmath,amssymb,nofootinbib]{revtex4}

\usepackage{epsfig}
\usepackage{graphicx}
\usepackage{dcolumn}
\usepackage{bm}

\def\DESepsf(#1 width #2){\epsfxsize=#2 \epsfbox{#1}}

\newcommand{\out}{{\rm out}}
\newcommand{\be}{\begin{eqnarray}}
\newcommand{\en}{\end{eqnarray}}
\newcommand{\ov}{\overline}
\newcommand{\A}{{\cal A}}
\newcommand{\B}{{\cal B}}
\newcommand{\Sc}{{\cal S}}
\newcommand{\T}{{\cal T}}
\newcommand{\U}{{\cal U}}
\newcommand{\V}{{\cal V}}

\begin{document}

\preprint{September, 2008}

\title{Rescattering effects in charmless $\overline B_{u,d,s}\to P P$ decays}
%

\author{Chun-Khiang Chua}
\affiliation{ Physics Department, Chung Yuan Christian University,
Chung-Li, Taiwan 32023, Republic of China
}%


\begin{abstract}
We study the final-state interaction (FSI) effects in charmless $\ov B_{u,d,s}\to PP$
decays. We consider a FSI approach with both short- and long-distance contributions,
where the former are from in-elastic
channels and are contained in factorization amplitudes, while the
latter are from the residual rescattering among $PP$ states. Flavor
SU(3) symmetry is used to constrain the residual rescattering
$S$-matrix. We fit to all available data on the CP-averaged decay
rates and CP asymmetries, and make predictions on unmeasured ones.
We investigate the $K\pi$ direct CP violations that lead to the
so-called $K\pi$ puzzle in CP violation. Our main results are
as follows: (i) Results are in agreement with data in the presence
of FSI. (ii) For $\ov B$ decays, the $\pi^+\pi^-$ and
$\pi^0\pi^0$ rates are suppressed and enhanced respectively by FSI.
(iii) The FSI has a large impact on direct CP
asymmetries ($\A$) of many modes. (iv) The deviation($\Delta \A$)
between $\A(\ov B{}^0\to K^-\pi^+)$ and $\A(B^-\to K^-\pi^0)$ can
be understood in the FSI approach. Since $\A(K^-\pi^0)$ is more
sensitive to the residual rescattering, the degeneracy of these
two direct CP violations can be successfully lifted. (v) Sizable
and complex color-suppressed tree amplitudes, which are crucial
for the large $\pi^0\pi^0$ rate and $\Delta\A$, are generated
through exchange rescattering. The correlation of the ratio
$\B(\pi^0\pi^0)/\B(\pi^+\pi^-)$ and $\Delta \A$ is studied.
(vi)~The $B^-\to \pi^-\pi^0$ direct CP violation is very small and
is not affected by FSI. (vii)~Several $\ov B_s$ decay rates are
enhanced. In particular, the $\eta'\eta'$ branching ratio is enhanced
to the level of $1.0\times 10^{-4}$, which can be checked experimentally.
(viii)~Time-dependent CP asymmetries $S$ in $\ov B_{d,s}$ decays
are studied. The $\Delta S(\ov B {}^0\to K_S\eta')$ is very small
($\leq 1\%$). This asymmetry remains to be one of the cleanest
measurements to search for new physics phases. The asymmetry $S$
from $\ov B_s$ to $PP$ states with strangeness S$=+1$ are expected
to be small. We found that the $|S|$ for $\ov B {}^0_s\to
\eta\eta$, $\eta\eta'$ and $\eta'\eta'$ decays are all below
$0.06$. CP asymmetries in these modes will be useful to test the
SM.
 \end{abstract}


\maketitle

\section{Introduction}

The study of $B$ decays provides many useful information of the
flavor sector of the Standard Model (SM)~\cite{PDG}. In
particular, the measurements of the time-dependent CP asymmetries
in kaon and charmonium final states give a rather precise value
of $\sin2\beta=0.681\pm0.025$~\cite{HFAG}, where
$\beta/\phi_1=\arg(V^*_{td})$ with $V$ the
Cabbibo-Kobayashi-Mashikawa (CKM) matrix. In the SM,
time-dependent CP asymmetries in penguin dominated modes are
expected to be close to the $\sin2\beta$ value~\cite{sin2beta}.
Since the penguin loop amplitudes are sensitive to high
virtuality, new physics beyond the SM may contribute to the time
dependent CP asymmetries through the heavy particles in the loops.
Consequently, these asymmetries are promising places to search for
new physics
effects~\cite{sin2beta,Cheng:2005bg,sin2betanew,Chua:2006hr,Gronau:2006qh,review}.

The measurements of direct CP violation ($\A$) in $\ov B$ decays
are also very useful and interesting. The $\A(\ov B{}^0\to
K^-\pi^+)$ asymmetry was the first measured direct CP violation in
$\ov B$ decays. The data confirmed a large $\A(\ov B{}^0\to
K^-\pi^+)$ with a negative sign as predicted in perturbative QCD
(pQCD) calculations~\cite{pQCD}. On the contrary, although
$\A(B^-\to K^-\pi^0)\simeq\A(\ov B{}^0\to K^-\pi^+)$ was expected
in many early theoretical predictions~\cite{pQCD,Beneke:2001ev},
the experimental evidence has been accumulated favoring a positive
$\A(B^-\to K^-\pi^0)$. The recent measurements show
$\A(K^-\pi^+)=(-9.8^{+1.2}_{-1.1})\%$ and
$\A(K^-\pi^0)=(5.0\pm2.5)\%$~\cite{HFAG}, giving $\Delta
\A(K\pi)\equiv \A(K^-\pi^0)-\A(K^-\pi^+)=(14.8^{+2.7}_{-2.8})\%$,
which is more than $5\,\sigma$ from zero. This is the so-called
$K\pi$ puzzle in direct CP violation, which has attracted a lot of
attentions~\cite{Buras:2004th,Buras:2004ub,Buras:2005cv,Li:2005kt,CharmingP,Baek2005,Hou:2006jy,Kim:2007kx,Nature}.
Several suggestions were put forward to resolve this puzzle. For
example, some authors introduced next-to-leading order
contributions in factorization amplitudes \cite{Li:2005kt}, while
some suggested new physics origins
\cite{Buras:2004th,Buras:2004ub,Buras:2005cv,Baek2005,Hou:2006jy,Kim:2007kx,Nature}
for the deviation.

It is well known that we need both weak and strong phase
differences to have a non-vanishing direct CP violation. Strictly
speaking the final state interaction (FSI) is the only source for
non-vanishing strong phases. In addition, it is capable
of enhancing the decay rates of many modes, which are measured to be larger than
expected. For example, the large observed $\ov
B{}^0\to\pi^0\pi^0$ rate, which remains puzzling and is still
posing tension in many theoretical calculations, can be obtained
by using FSI~\cite{Hou:1999st}. Furthermore, it was realized
recently that long-distance FSI may play an indispensable role in
charmful as well as in charmless $\ov B$
decays~\cite{Chua:2001br,Cheng:2004ru}.

Data for $\overline B_s$ decays are starting to emerge from the
Tevetron~\cite{PDG} and from $B$ factories, and we anticipate more
to come in the near future, from LHCb and other LHC experiments.
Measurements of rates and CP asymmetries in $\ov B_s$ decays will
be useful in testing the SM and in searching for new (physics)
phases. In fact, recently, a claim on the evidence of new physics
effect in the $\ov B{}_s$ mixing was put forward~\cite{UTfit}.  It is
thus timely to study $\overline B_s$ decays.

In this work, we investigate the effects of FSI on all charmless
$\ov B_{u,d,s}\to PP$ decay rates and CP asymmetries. We outline
the underlying physical picture of the FSI approach employed in
this work.
The master formula of FSI for charmless $\ov B\to PP$ decays is
(see appendix A, if a derivation is needed)
 \be
 A^{FSI}_i=\sum_{k=1}^N\Sc^{1/2}_{ik} A^0_k,
 \label{eq:master}
 \en
where $A^{FSI}$ and $A^0$ are $\ov B$ decay amplitudes with and
without FSI~\footnote{Note that $A^{FSI}$ contains weak as well as
strong phases, while $A^0$ only has weak phases.}, respectively,
$i=1,\dots,n$, denotes all charmless $PP$ states,
$k=1,\dots,n,n+1,\dots,N,$ denotes {\it all} possible states that
can rescatter into the charmless $PP$ states and $\Sc$ is the
strong interacting $S$-matrix. Note that no approximation has been
made in the above equation, which, in principle, all charmless
$\ov B$ decay amplitudes should follow. In practice, this master
formula is hard to use as it involves many states (the number $N$
is in general quite large in a typical charmless $B$ decay).

Let us investigate further the difficulties of using the above
master formula. The number of states allowed to enter the formula
grows with the mass of the decaying particle. For a typical $B$
decay, there is a large number of the states involved in the equation  and
the contributions are hard to handle. For example, we may need to
consider a rescattering process contributed from a multi-body final state,
where the decay amplitude and the corresponding rescattering
$S$-matrix element are poorly known. Therefore, the complication
originates from the largeness of $m_B$. However, it is precisely
the largeness of $m_B$ that makes factorization approaches such
as pQCD~\cite{pQCD}, QCD factorization
(QCDF)~\cite{Beneke:2001ev,Beneke:2003zv} and soft collinear
effective theory (SCET)~\cite{SCET} possible. These approaches
achieve accessibility and simplifications. The underlying reason
for the simplicity is related to the so-called duality argument,
which uses the fact that when contributions from all hadronic
states at a large enough energy scale are summed over, one should
be able to understand the physics in terms of the quark and gluon
degrees of freedom. Hence, it is reasonable to expect that the
main effect of FSI, especially those from inelastic channels, is
included in the factorization amplitudes -- a statement we expect
to hold perfectly in the $m_b\to\infty$ case. Since in the real
world $m_b$ is finite, whether it is large enough to validate the
above argument should be answered by experiments.

It is fair to say that most factorization results on CP-averaged
charmless $\ov B\to PP$ decay rates, especially color-allowed
ones, agree well with data. However, some measurements seem to
imply the needs of sub-leading contributions. For example, rates
of some suppressed decay modes, such as the above mentioned $\ov
B{}^0\to\pi^0\pi^0$ rate, and some CP-odd quantities, such as the
$B^-\to K^-\pi^0$ direct CP violation, do not agree well with
predictions. These are places, where sub-leading effects, such as
FSI, could be visible. Therefore, although we expect factorization
amplitudes to contain most of the FSI effects demanded in
Eq.~(\ref{eq:master}), it is likely that residual rescattering is
still allowed and needed in $\ov B\to PP$ decays at the physical
$m_B$ energy scale. The group of charmless $PP$ states is unique
to the processes we are studying and is well separated from all
other states. Since the duality argument cannot be applied to
these states of limited number, part of their FSI effects may not
be included in the factorization
amplitudes~\cite{quasielastic0,quasielastic}.

In summary, FSI in $B$ decays may be simpler than we thought,
since $m_B$ could be large enough to apply a
factorization approach for the main part of FSI contributions. We
may only need to include the left-over FSI, i.e. residual
rescattering, in addition to the short-distance FSI in the $PP$
sector. In this sense, the FSI approach we are using is a mild
extension to the factorization approaches.

Note that a similar approach analyzing early data was used in
\cite{quasielastic}. There is one major difference. In
\cite{quasielastic}, in principle, no short-distance phase was
allowed in factorization amplitudes to avoid double counting,
while here we do need short-distance phases to account for the FSI
effects from all in-elastic plus some quasi-elastic channels.
There are
also other works in the literature discussing rescattering among
$PP$ states and/or from some in-elastic
channels~\cite{Cheng:2004ru,Smith,Donoghue:1996hz,Zenczykowski:2003ed,Suzuki:2007je,Wu:2007tm}.
For example, in \cite{Cheng:2004ru}, rescattering from $PP$ and
$D^{(*)}\bar D^{(*)}$ final states was considered, and the main
FSI contributions resemble the charming penguin
ones~\cite{RGI,Wu:2007tm}. We also note that similar discussion of
the factorization of $\Sc$ into short-distance and residual parts,
as well as the discussion of the approximation done when assuming
$S$ is block diagonal (with a block for the $PP$ states) can be
found in \cite{Smith}.

The layout of the present paper is as follows: In Sec. II we
introduce the formalism.  We then use it in Sec. III to study $\ov
B_{u,d,s}\to PP$ decays. Results and discussions are presented in
Sec. IV. Sec. V contains our conclusions. Some derivations,
including those lead to Eq.~(\ref{eq:master}), are given in
Appendices.

\section{Formalism}

In this section, we develop the formalism. The reader who is not interested
in the detail of the formalism may proceed directly
to 
the numerical analysis section.

Without loss of generality, we can re-express the $S$-matrix in
Eq.~(\ref{eq:master}) as
 \be
 \Sc_{ik}=\sum_{j=1}^n(\Sc_1)_{ij} (\Sc_2)_{jk},
 \label{eq:S1S2}
 \en
where $\Sc_1$ is a non-singular $n\times n$ matrix with $n$ the
total number of charmless $PP$ states and $\Sc_2$ is defined
through the above equation, i.e. $\Sc_2\equiv \Sc^{-1}_{1} \Sc$.
The physical picture mentioned in the last section is close to the
one in factorization approaches, except that there are still some
residual rescattering effects, and we have
 \be
 \Sc_1=\Sc_{res},\quad
 A^{fac}_j=\sum_{k=1}^N(\Sc^{1/2}_2)_{jk}A^0_k,
 \label{eq:res}
 \en
with $N$ the total number of states entering Eq.~(\ref{eq:master})
and $A^{fac}_j$ the factorization amplitude. The residual
rescattering effect is encoded in the $\Sc_{res}$ matrix.  Note
that although $\Sc$ is unitary, $\Sc_{res}$ needs not be so, as it
describes the residual rescattering among various charmless $PP$
states. In factorization approaches, the above $\Sc_{res}$ is
taken to be unity. We shall use the up-to-date data to
determine $\Sc_{res}$. It should be reminded that our framework
does not exclude the fully factorized case ($\Sc_{res}=1$) and,
hence, it is also being tested.
To apply the above formula, we need to specify the factorization
amplitudes. In this work, we use the factorization amplitudes
obtained in the QCD factorization approach~\cite{Beneke:2003zv}.

Combining Eqs.~(\ref{eq:master}) and (\ref{eq:res}), we have
 \be
 A^{FSI}_i=\sum_{j=1}^n(\Sc_{res}^{1/2})_{ij} A^{fac}_j,
 \label{eq:master1}
 \en
where, as mentioned before, $i,j=1,\dots,n$ denote all charmless
$PP$ states. The number of parameters needed to describe
$\Sc_{res}$ seems to be quite large. This is, however, not the
case, since strong interaction has (an approximate) SU(3)
symmetry, which is expected to be a good one at the $m_B$
rescattering scale and, hence, can be used to constrain the form
of $S_{res}$.

\begin{figure}[t!]
\centerline{\DESepsf(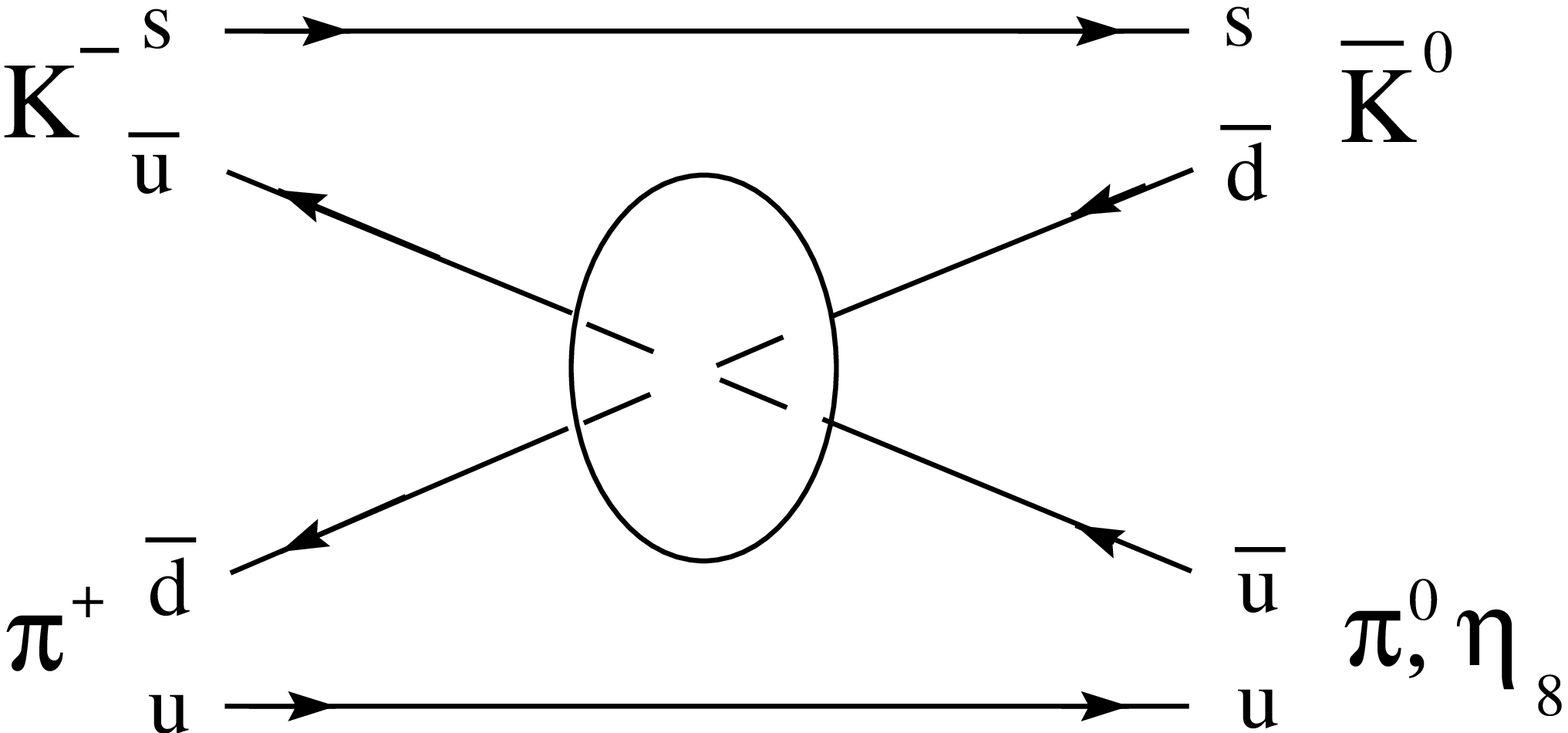 width 6cm)
            \hspace{0.6cm}
            \DESepsf(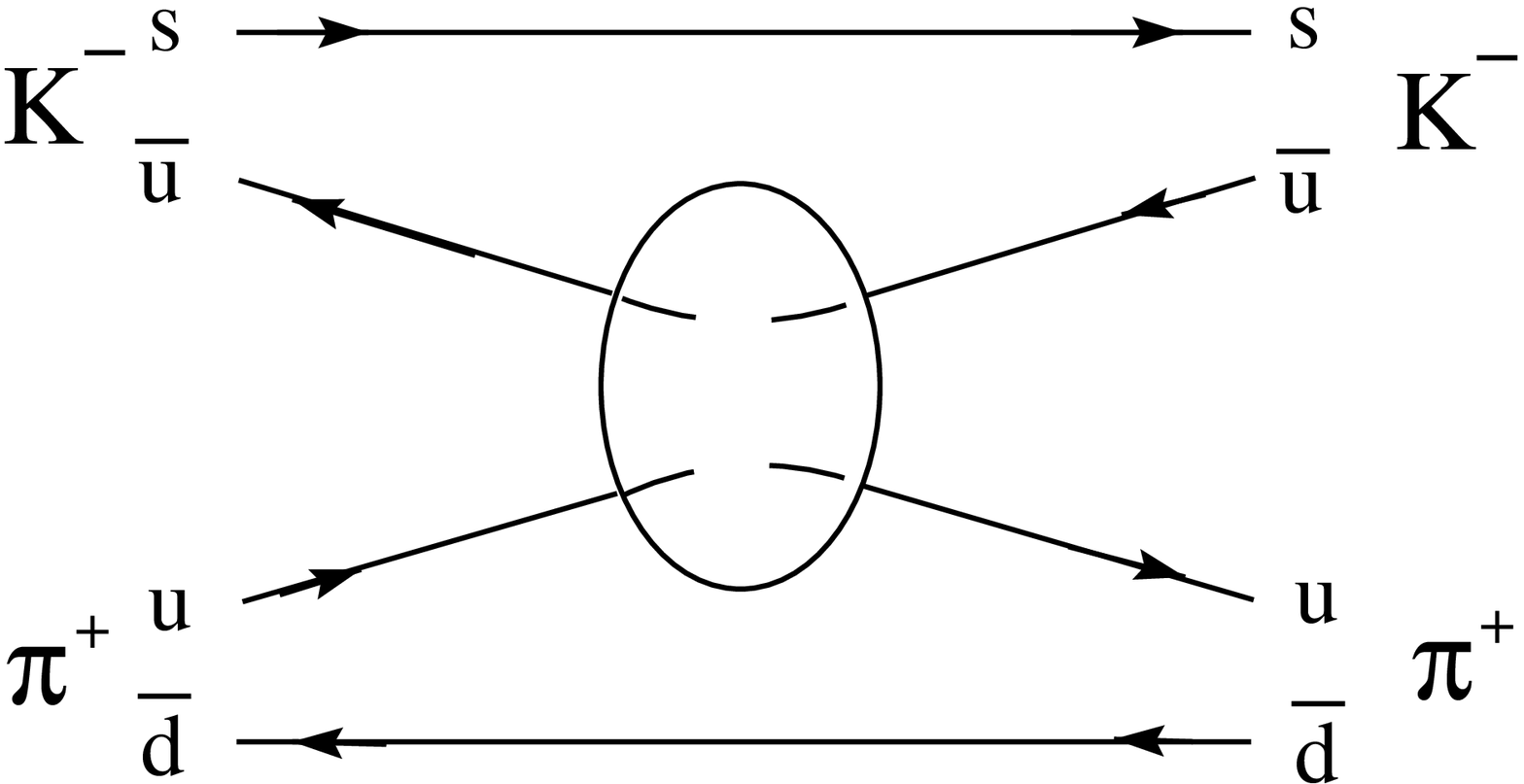 width 5.7cm)}
\centerline{(a)\hspace{6.5cm}(b)}
\vspace{0.3cm}
\centerline{\DESepsf(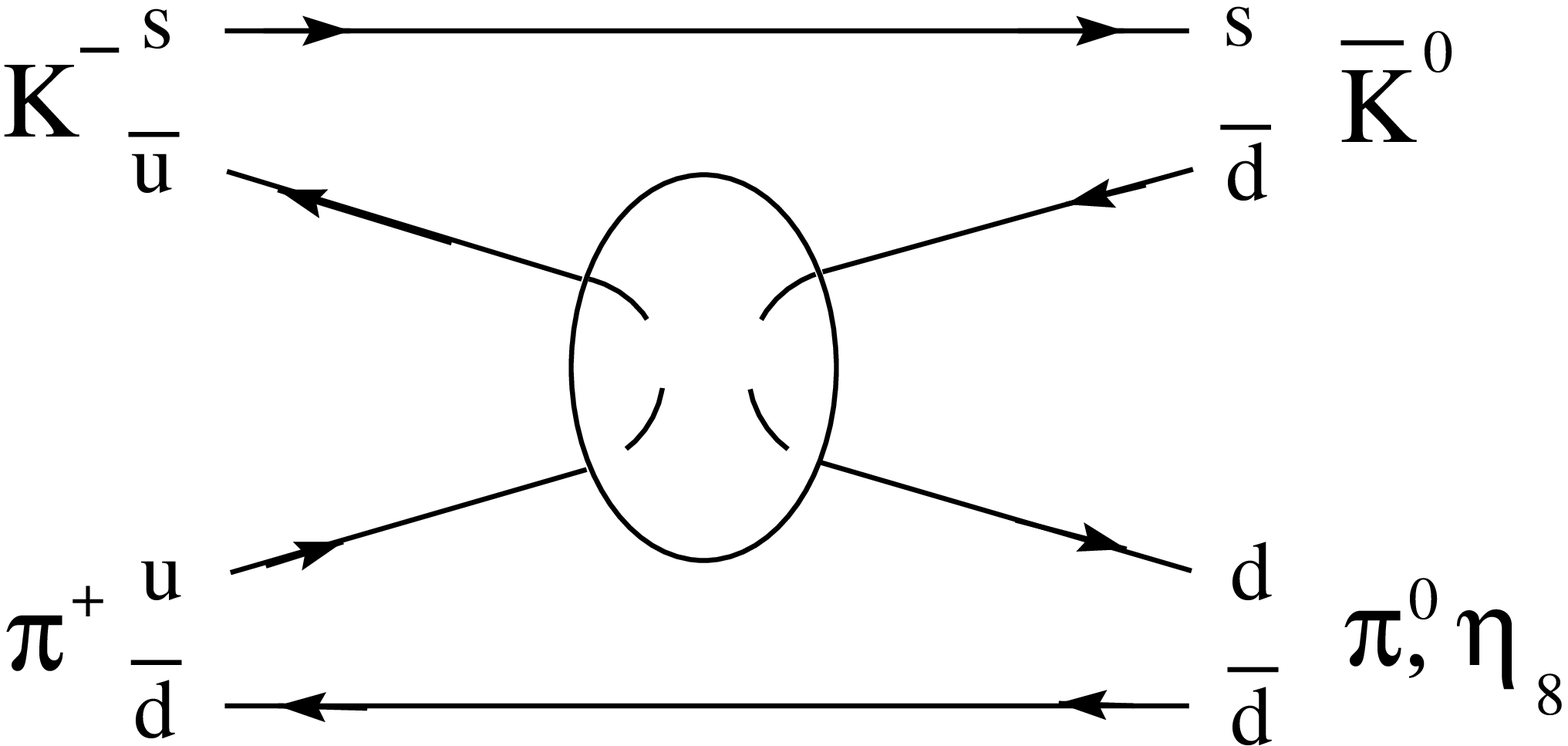 width 6cm)
            \hspace{0.6cm}
            \DESepsf(ra.eps width 6cm)}
\centerline{(c)\hspace{6.5cm}(d)}
\caption{Pictorial representation of
   (a) charge exchange $r_e$, (b) singlet exchange $r_0$,
   (c) annihilation $r_a$ and (d) total-annihilation $r_t$
   for $PP$ (re)scattering.} \label{fig:r}

\end{figure}

Explicitly, through SU(3) symmetry, we have
\begin{eqnarray}
\left(
\begin{array}{l}
 A^{FSI}_{\ov B {}^0_{d,s}\to K^-\pi^+}\\
 A^{FSI}_{\ov B {}^0_{d,s}\to \ov K {}^0 \pi^0}\\
 A^{FSI}_{\ov B {}^0_{d,s}\to \ov K {}^0 \eta_8}\\
 A^{FSI}_{\ov B {}^0_{d,s}\to \ov K {}^0 \eta_1}
\end{array}
\right)
 &=& \Sc_{res,1}^{1/2}
 \left(
\begin{array}{l}
 A^{fac}_{\ov B {}^0_{d,s}\to K^-\pi^+}\\
 A^{fac}_{\ov B {}^0_{d,s}\to \ov K {}^0 \pi^0}\\
 A^{fac}_{\ov B {}^0_{d,s}\to \ov K {}^0 \eta_8}\\
 A^{fac}_{\ov B {}^0_{d,s}\to \ov K {}^0 \eta_1}
\end{array}
\right),
 \label{eq:FSIB0Kpi}
\end{eqnarray}
\begin{eqnarray}
\left(
\begin{array}{l}
 A^{FSI}_{B^-\to \ov K^0\pi^-}\\
 A^{FSI}_{B^-\to K^- \pi^0}\\
 A^{FSI}_{B^-\to K^- \eta_8}\\
 A^{FSI}_{B^-\to K^- \eta_1}
\end{array}
\right)
 &=& \Sc_{res,2}^{1/2}
 \left(
\begin{array}{l}
 A^{fac}_{B^-\to \ov K^0\pi^-}\\
 A^{fac}_{B^-\to K^- \pi^0}\\
 A^{fac}_{B^-\to K^- \eta_8}\\
 A^{fac}_{B^-\to K^- \eta_1}
\end{array}
\right),
 \label{eq:FSIBKpi0}
\end{eqnarray}
\begin{eqnarray}
\left(
\begin{array}{l}
 A^{FSI}_{B^-\to \pi^-\pi^0}\\
 A^{FSI}_{B^-\to K^0 K^-}\\
 A^{FSI}_{B^-\to \pi^- \eta_8}\\
 A^{FSI}_{B^-\to \pi^- \eta_1}
\end{array}
\right)
 &=& \Sc_{res,3}^{1/2}
 \left(
\begin{array}{l}
 A^{fac}_{B^-\to \pi^-\pi^0}\\
 A^{fac}_{B^-\to K^0 K^-}\\
 A^{fac}_{B^-\to \pi^- \eta_8}\\
 A^{fac}_{B^-\to \pi^- \eta_1}
\end{array}
\right),
 \label{eq:FSIBpipi0}
\end{eqnarray}
\begin{eqnarray}
\left(
\begin{array}{l}
 A^{FSI}_{\ov B {}^0_{d,s}\to \pi^+\pi^-}\\
 A^{FSI}_{\ov B {}^0_{d,s}\to \pi^0 \pi^0}\\
 A^{FSI}_{\ov B {}^0_{d,s}\to \eta_8 \eta_8}\\
 A^{FSI}_{\ov B {}^0_{d,s}\to \eta_8 \eta_1}\\
 A^{FSI}_{\ov B {}^0_{d,s}\to \eta_1 \eta_1}\\
 A^{FSI}_{\ov B {}^0_{d,s}\to K^+ K^-}\\
 A^{FSI}_{\ov B {}^0_{d,s}\to K^0 \ov K {}^0}\\
 A^{FSI}_{\ov B {}^0_{d,s}\to \pi^0 \eta_8}\\
 A^{FSI}_{\ov B {}^0_{d,s}\to \pi^0 \eta_1}
\end{array}
\right)
 &=& \Sc_{res,4}^{1/2}
 \left(
\begin{array}{l}
 A^{fac}_{\ov B {}^0_{d,s}\to \pi^+\pi^-}\\
 A^{fac}_{\ov B {}^0_{d,s}\to \pi^0 \pi^0}\\
 A^{fac}_{\ov B {}^0_{d,s}\to \eta_8 \eta_8}\\
 A^{fac}_{\ov B {}^0_{d,s}\to \eta_8 \eta_1}\\
 A^{fac}_{\ov B {}^0_{d,s}\to \eta_1 \eta_1}\\
 A^{fac}_{\ov B {}^0_{d,s}\to K^+ K^-}\\
 A^{fac}_{\ov B {}^0_{d,s}\to K^0 \ov K {}^0}\\
 A^{fac}_{\ov B {}^0_{d,s}\to \pi^0 \eta_8}\\
 A^{fac}_{\ov B {}^0_{d,s}\to \pi^0 \eta_1}
\end{array}
\right),
 \label{eq:FSIB0pipi}
\end{eqnarray}
%
where we have ${\cal S}^{1/2}_{res,i}=(1+i{\cal T}_i)^{1/2}$, with
\begin{eqnarray}
 {\cal T}_1 &=& \left(
\begin{array}{cccc}
r_0+r_a
      &\frac{-r_a+r_e}{\sqrt2}
      &\frac{-r_a+r_e}{\sqrt6}
      &\frac{2\bar r_a+\bar r_e}{\sqrt3}
      \\
\frac{-r_a+r_e}{\sqrt2}
      &r_0+\frac{r_a+r_e}{2}
      &\frac{r_a-r_e}{2\sqrt3}
      &-\frac{2\bar r_a+\bar r_e}{3\sqrt2}
      \\
\frac{-r_a+r_e}{\sqrt6}
      &\frac{r_a-r_e}{2\sqrt3}
      &r_0+\frac{r_a+5r_e}{6}
      &-\frac{2\bar r_a+\bar r_e}{3\sqrt2}
      \\
\frac{2\bar r_a+\bar r_e}{\sqrt3}
      &-\frac{2\bar r_a+\bar r_e}{\sqrt6}
      &-\frac{2\bar r_a+\bar r_e}{3\sqrt2}
      &\tilde r_0+\frac{4\tilde r_a+2\tilde r_e}{3}
\end{array}
\right), \nonumber\\
 {\cal T}_2 &=& \left(
\begin{array}{cccc}
r_0+r_a
      &\frac{r_a-r_e}{\sqrt2}
      &\frac{-r_a+r_e}{\sqrt6}
      &\frac{2\bar r_a+\bar r_e}{\sqrt3}
      \\
\frac{r_a-r_e}{\sqrt2}
      &r_0+\frac{r_a+r_e}{2}
      &\frac{-r_a+r_e}{2\sqrt3}
      &\frac{2\bar r_a+\bar r_e}{3\sqrt2}
      \\
\frac{-r_a+r_e}{\sqrt6}
      &\frac{-r_a+r_e}{2\sqrt3}
      &r_0+\frac{r_a+5r_e}{6}
      &-\frac{2\bar r_a+\bar r_e}{3\sqrt2}
      \\
\frac{2\bar r_a+\bar r_e}{\sqrt3}
      &\frac{2\bar r_a+\bar r_e}{\sqrt6}
      &-\frac{2\bar r_a+\bar r_e}{3\sqrt2}
      &\tilde r_0+\frac{4\tilde r_a+2\tilde r_e}{3}
\end{array}
\right), \nonumber\\
 {\cal T}_3 &=& \left(
\begin{array}{cccc}
r_0+r_a
      &0
      &0
      &0
      \\
0
      &r_0+r_a
      &\sqrt{\frac{2}{3}}(r_a-r_e)
      &\frac{2\bar r_a+\bar r_e}{\sqrt3}
      \\
0
      &\sqrt{\frac{2}{3}}(r_a-r_e)
      &r_0+\frac{2r_a+r_e}{3}
      &\frac{\sqrt2}{3}(2\bar r_a+\bar r_e)
      \\
0
      &\frac{2\bar r_a+\bar r_e}{\sqrt3}
      &\frac{\sqrt2}{3}(2\bar r_a+\bar r_e)
      &\tilde r_0+\frac{4\tilde r_a+2\tilde r_e}{3}
\end{array}
\right),
 \label{eq:T123}
\end{eqnarray}
and
\begin{eqnarray}
&&\hspace{-1.5cm}\T_4={\rm diag}(r_0,r_0,r_0,\tilde r_0,\check
r_0,r_0,r_0,r_0,\tilde r_0)
\nonumber\\
 &&\hspace{-1.5cm}+ \small{
 \left(
\begin{array}{ccccccccc}
2r_a+r_t
       &\frac{2r_a-r_e+r_t}{\sqrt2}
       &\frac{2r_a+r_e+3r_t}{3\sqrt2}
       &\frac{\sqrt2(2\bar r_a+\bar r_e)}{3}
       &\frac{4\hat r_a+2\hat r_e+3\hat r_t}{3\sqrt2}
       &r_a+r_t
       &r_a+r_t
       &0
       &0
       \\
\frac{2r_a-r_e+r_t}{\sqrt2}
       &\frac{2r_a+r_e+r_t}{2}
       &\frac{2r_a+r_e+3r_t}{6}
       &\frac{2\bar r_a+\bar r_e}{3}
       &\frac{4\hat r_a+2\hat r_e+3\hat r_t}{6}
       &\frac{r_a+r_t}{\sqrt2}
       &\frac{r_a+r_t}{\sqrt2}
       &0
       &0
       \\
\frac{2r_a+r_e+3r_t}{3\sqrt2}
       &\frac{2r_a+r_e+3r_t}{6}
       &\frac{2r_a+r_e+r_t}{2}
       &-\frac{2\bar r_a+\bar r_e}{3}
       &\frac{4\hat r_a+2\hat r_e+3\hat r_t}{6}
       &\frac{5r_a-2r_e+3r_t}{3\sqrt2}
       &\frac{5r_a-2r_e+3r_t}{3\sqrt2}
       &0
       &0
       \\
\frac{\sqrt2(2\bar r_a+\bar r_e)}{3}
       &\frac{2\bar r_a+\bar r_e}{3}
       &-\frac{2\bar r_a+\bar r_e}{3}
       &\frac{4\tilde r_a+2\tilde r_e}{3}
       &0
       &-\frac{2\bar r_a+\bar r_e}{3\sqrt2}
       &-\frac{2\bar r_a+\bar r_e}{3\sqrt2}
       &0
       &0
       \\
\frac{4\hat r_a+2\hat r_e+3\hat r_t}{3\sqrt2}
       &\frac{4\hat r_a+2\hat r_e+3\hat r_t}{6}
       &\frac{4\hat r_a+2\hat r_e+3\hat r_t}{6}
       &0
       &\frac{4\check r_a+2\check r_e+3\check r_t}{6}
       &\frac{4\hat r_a+2\hat r_e+3\hat r_t}{3\sqrt2}
       &\frac{4\hat r_a+2\hat r_e+3\hat r_t}{3\sqrt2}
       &0
       &0
       \\
r_a+r_t
       &\frac{r_a+r_t}{\sqrt2}
       &\frac{5r_a-2r_e+3r_t}{3\sqrt2}
       &-\frac{2\bar r_a+\bar r_e}{3\sqrt2}
       &\frac{4\hat r_a+2\hat r_e+3\hat r_t}{3\sqrt2}
       &2 r_a+r_t
       &r_a+r_t
       &\frac{r_a-r_e}{\sqrt3}
       &\frac{2\bar r_a+\bar r_e}{\sqrt6}
       \\
r_a+r_t
       &\frac{r_a+r_t}{\sqrt2}
       &\frac{5r_a-2r_e+3r_t}{3\sqrt2}
       &-\frac{2\bar r_a+\bar r_e}{3\sqrt2}
       &\frac{4\hat r_a+2\hat r_e+3\hat r_t}{3\sqrt2}
       &r_a+r_t
       &2r_a+r_t
       &\frac{-r_a+r_e}{\sqrt3}
       &-\frac{2\bar r_a+\bar r_e}{\sqrt6}
       \\
0
       &0
       &0
       &0
       &0
       &\frac{r_a-r_e}{\sqrt3}
       &\frac{-r_a+r_e}{\sqrt3}
       &\frac{2r_a+r_e}{3}
       &\frac{\sqrt2(2\bar r_a+\bar r_e)}{3}
       \\
0
       &0
       &0
       &0
       &0
       &\frac{2\bar r_a+\bar r_e}{\sqrt6}
       &-\frac{2\bar r_a+\bar r_e}{\sqrt6}
       &\frac{\sqrt2(2\bar r_a+\bar r_e)}{3}
       &\frac{4\tilde r_a+2\tilde r_e}{3}
       \end{array}
 \right)
 }.
 \nonumber\\
\label{eq:T4}
\end{eqnarray}
The rescattering parameters $r_{0,a,e,t}$, $\bar r_{0,a,e,t}$,
$\tilde r_{0,a,e,t}$, $\hat r_{0,a,e,t}$ and $\check r_{0,a,e,t}$
denote rescattering in $\Pi({\bf 8})\,\Pi({\bf 8})\to\Pi({\bf
8})\,\Pi({\bf 8})$, $\Pi({\bf 8})\,\Pi({\bf 8})\to\Pi({\bf
8})\,\eta_1$, $\Pi({\bf 8})\eta_1\to\Pi({\bf 8})\eta_1$ and
$\eta_1\eta_1\to\eta_1\eta_1$, respectively, and the subscripts
$0,a,e,t$ represent flavor singlet, annihilation, exchange and
total-annihilation rescatterings, respectively (see
Fig.~\ref{fig:r}). Note that for identical particle final states,
such as $\pi^0\pi^0$, factors of $1/\sqrt2$ are included in the
amplitudes and the corresponding $\Sc_{res}$ matrix elements. The
$P\eta_8,P\eta_1$ are not physical final states. The physical
$\eta,\,\eta^\prime$ mesons are defined through
\begin{equation}
\left(
\begin{array}{c}
\eta\\
      \eta^\prime
\end{array}
\right)= \left(
\begin{array}{cc}
\cos\vartheta &-\sin\vartheta\\
\sin\vartheta &\cos\vartheta
\end{array}
\right) \left(
\begin{array}{c}

\eta_8\\
      \eta_1
\end{array}
\right),
\end{equation}
with the mixing angle $\vartheta\simeq-15.4^\circ$
\cite{Feldmann:1998vh}. For the $\eta^{(\prime)}\eta^{(\prime)}$
states, we have
\begin{equation}
\left(
\begin{array}{c}
\eta\eta\\
      \eta\eta^\prime\\
      \eta'\eta'
\end{array}
\right)= \left(
\begin{array}{ccc}
\cos^2\vartheta
   &-\sqrt2\cos\vartheta \sin\vartheta
   &\sin^2\vartheta\\
\sqrt2\cos\vartheta \sin\vartheta
   &\cos^2\vartheta-\sin^2\vartheta
   &-\sqrt2\cos\vartheta \sin\vartheta\\
\sin^2\vartheta
   &\sqrt2\cos\vartheta \sin\vartheta
   &\cos^2\vartheta
\end{array}
\right) \left(
\begin{array}{c}
\eta_8\eta_8\\
      \eta_8\eta_1\\
      \eta_1\eta_1
\end{array}
\right),
\end{equation}
where the identical particle factor of $1/\sqrt2$ is properly
included in the mixing matrix.

The matrices $\T_{1,2,3,4}$ can be obtained through a diagrammatic
method by matching the Clebsh-Gordan coefficients of scattering
mesons (see Fig.~\ref{fig:r}) or by using an operator method. We
have $Tr(\Pi_1^{\rm in}\Pi_1^{\rm out}\Pi_2^{\rm in}\Pi_2^{\rm
out})/2$, $Tr(\Pi_1^{\rm in}\Pi_2^{\rm in}\Pi_1^{\rm
out}\Pi_2^{\rm out})$, $Tr(\Pi_1^{\rm in}\Pi_1^{\rm
out})Tr(\Pi_2^{\rm in}\Pi_2^{\rm out})$ and $Tr(\Pi_1^{\rm
in}\Pi_2^{\rm in})Tr(\Pi_1^{\rm out}\Pi_2^{\rm out})$
corresponding to $r_e$, $r_a$, $r_0$ and $r_t$ contributions,
respectively (see similar discussion for the case of charmful $B$
decays in \cite{Chua:2001br}). Note that due to Bose-Einstein
statistic and the $S$-wave configuration of the final state mesons
in $\ov B\to P_1P_2$ decays, the amplitude should be symmetric
under the exchange of the indices $1$ and $2$. Consequently, the
above terms exhaust all possible combinations for $\Pi({\bf
8})\,\Pi({\bf 8})\to\Pi({\bf 8})\,\Pi({\bf 8})$ scatterings. For
operators involving $\eta_1$, we suitably replace $\Pi$ in the
above expressions by $\eta_1 1_{3\times 3}$ to obtain operators
corresponding to $\bar r_i$, $\tilde r_i$, $\hat r_i$ and $\check
r_i$.

It can be easily seen that rescattering formulas for charmless
$\ov B_s\to PP$ decays resemble those for $\ov B {}^0\to PP$
decays. Information on $\Sc_{rec}$ obtained from $\ov B {}^0_d$
decays can be used to predict $\ov B_s$ decay rates.

At first sight, it appears that we need 40 real parameters (from
20 complex rescattering parameters: $r_{0,a,e,t}$, $\bar
r_{0,a,e,t}$, $\tilde r_{0,a,e,t}$, $\hat r_{0,a,e,t}$ and $\check
r_{0,a,e,t}$) to describe $\Sc_{res}$. The number of independent parameters is
actually much lower for two reasons. First, rescattering
parameters enter $\Sc_{res}$ only through 7 independent
combinations: $1+i(r_0+r_a)$, $i(r_e-r_a)$, $i(r_a+r_t)$, $i(2\bar
r_a+\bar r_e)$, $1+i[\tilde r_0+(4\tilde r_a+2\tilde r_e)/3]$,
$i(4\hat r_a+2\hat r_e+3\hat r_t)$ and $1+i[\check r_0+(4\check
r_a+2\check r_e+3\check r_t)/6]$. Second, SU(3) symmetry imposes
further constraints on these combinations.

Flavor symmetry requires that $(\Sc_{res})^m$ with an arbitrary
power of $m$ should also have the same form as $\Sc_{res}$. More
explicitly, from SU(3) symmetry, we should have
 \be
 (\Sc_{res})^m&=&(1+i\T)^m\equiv 1+i\T^{(m)},
 \label{eq:SSU3a}
 \en
where $\T^{(m)}$ is defined through the above equation and its
form is given by
 \be
 \T^{(m)}&=&\T\,\,\, {\rm with\,\,}(r_j,\bar r_j,\tilde r_j,\check r_j)
 \to (r^{(m)}_j,\bar r^{(m)}_j,\tilde r^{(m)}_j,\check r^{(m)}_j),
 \label{eq:SSU3b}
 \en
for $j=0,a,e,t$.

It is found that the solutions to Eqs.~(\ref{eq:SSU3a}) and
(\ref{eq:SSU3b}) are given by
 \be
 1+i(r^{(m)}_0+r^{(m)}_a)
      &=&\frac{2 e^{2m\,i\delta_{27}}+3\U^{m}_{11}}{5},
      \nonumber\\
 i(r^{(m)}_e-r^{(m)}_a)
      &=&\frac{3 e^{2m\,i\delta_{27}}-3\U^{m}_{11}}{5},
      \nonumber\\
 i (r^{(m)}_a+r^{(m)}_t)
      &=&\frac{-e^{2m\,i\delta_{27}}- 4 \U^{m}_{11} + 5 \V^{m}_{11}}{20},
      \nonumber\\
 i(2\bar r^{(m)}_a+\bar r^{(m)}_e)
      &=&\frac{3}{\sqrt5} \U^{m}_{12},
      \nonumber\\
 1+i(\tilde r^{(m)}_0+\frac{4\tilde r^{(m)}_a+2\tilde r_e^{(m)}}{3})
      &=&\U_{22}^{m},
      \nonumber\\
 i(4\hat r^{(m)}_a+2\hat r^{(m)}_e+3\hat r^{(m)}_t)
      &=&\frac{3}{\sqrt2} \V^{m}_{12},
      \nonumber\\
  1+i(\check r^{(m)}_0+\frac{4\check r^{(m)}_a+2\check r_e^{(m)}+3\check r_t}{6})
      &=&\V_{22}^{m},
 \label{eq:solution}
 \en
where $\U^{m}_{ij}$ and $\V^{m}_{ij}$ are elements of
 \be
\U^{m}(\tau,\delta_8,\delta'_{8})
 &\equiv&\left(
\begin{array}{cc}
\cos\tau
       &\sin\tau
       \\
-\sin\tau
       &\cos\tau
\end{array}
\right) \left(
\begin{array}{cc}
e^{2m\,i\delta_8}
       &0
       \\
0
       &e^{2m\,i\delta_8'}
\end{array}
\right) \left(
\begin{array}{cc}
\cos\tau
       &-\sin\tau
       \\
\sin\tau
       &\cos\tau
\end{array}
\right),
 \nonumber\\
 \V^{m}(\nu,\delta_1,\delta'_{1})
 &\equiv&\left(
\begin{array}{cc}
\cos\nu
       &\sin\nu
       \\
-\sin\nu
       &\cos\nu
\end{array}
\right) \left(
\begin{array}{cc}
e^{2m\,i\delta_1}
       &0
       \\
0
       &e^{2m\,i\delta_1'}
\end{array}
\right) \left(
\begin{array}{cc}
\cos\nu
       &-\sin\nu
       \\
\sin\nu
       &\cos\nu
\end{array}
\right),
 \label{eq:UandV}
 \en
respectively. From the above solution, we see that two real mixing
angles $\tau$ and $\nu$, and five {\it complex} phases
$\delta_{27,8,1},\delta'_{8,1}$ are needed to describe
$(\Sc_{res})^m$ in the full SU(3) case.

Several remarks are in order. (i) The subscripts of phases denote
the corresponding SU(3) multiplets and more details will be
given shortly.
(ii)~The imaginary parts of $\delta_{27,8,1},\delta'_{8,1}$
control the leakage of the non-unitary $\Sc^m_{res}$ through the
scattering of $PP$ states into non-$PP$ states.
(iii)~As we shall see, the $\Sc^m_{res}$ can be factorized into
two parts depending only on the real and the imaginary parts
of these phases, respectively.
(vi)~To reduce the number of the FSI parameters we will consider a
restricted SU(3) case, which is close to a U(3) symmetric case.

Since charmless mesons $P$ consist of an octet ($\Pi(\bf 8)$) and
an singlet ($\eta_1$), we have $\bf 8\otimes 8$, $\bf 8\otimes 1$,
$\bf 1\otimes 8$ and $\bf 1\otimes 1$ SU(3) products for $P_1P_2$
final states. Due to the $S$-wave configuration of $P_1P_2$ in
$\ov B$ decays and the Bose-Einstein statistics, the resulting
SU(3) multiplets should be symmetric under the exchange of $P_1$ and
$P_2$. The allowed ones are the $\bf 27$, $\bf 8$ and the $\bf 1$
from $\bf 8\otimes 8$, the $\bf 8'$ from the symmetrized $\bf
8\otimes 1$+$\bf 1\otimes 8$, and $\bf 1'$ from $\bf 1\otimes1$
(see, for example \cite{TDLee}). Hence, from SU(3) symmetry and
the Bose-Einstein statistics, we have
 \be
 (\Sc_{res})^m
 =\sum_{a=1}^{27}|{\bf 27};a\rangle e^{2m\,i\delta_{27}}\langle {\bf 27};a|
  +\sum_{b=1}^{8}\sum_{p,q=8,8'}|p;b\rangle \U^{m}_{pq} \langle q;b|
  +\sum_{p,q=1,1'}|p;1\rangle \V^{m}_{pq} \langle q;1|,
 \label{eq:SU3decomposition}
 \en
where $a$ and $b$ are labels of states within multiplets. It can
be easily seen that the above form of $\Sc^m_{res}$ is preserved
for any value of $m$.
We note that similar formulas for $\ov B\to PP$ rescattering
(excluding $P=\eta_1$) from SU(3) symmetry have been used in
\cite{quasielastic,Smith}.

From Eq.~(\ref{eq:SU3decomposition}) we see that the matrix
$\Sc_{res}^m$ is in block-diagonal form and we also have
 \be
 \U^{m}(\tau,\delta_8,\delta'_{8})=\U^{m}(\tau,{\rm Re}\,\delta_8,{\rm Re}\,\delta'_{8})\cdot\U^{m}(\tau,i{\rm
 Im}\,\delta_8,i{\rm Im}\,\delta'_{8}),
 \nonumber\\
 \V^{m}(\nu,\delta_1,\delta'_{1})=\V^{m}(\nu,{\rm Re}\,\delta_1,{\rm Re}\,\delta'_{1})\cdot\V^{m}(\nu,i{\rm Im}\,\delta_1,i{\rm
 Im}\,\delta'_{1}),
 \en
which can be proved by using the explicit expressions of $\U^m$
and $\V^m$ given in Eq. (\ref{eq:UandV}). Consequently, the matrix
$\Sc_{res}^m$ can be factorized into two matrices containing only
real and imaginary phases, respectively, i.e.
 \be
\Sc_{res}^m(\tau,\nu;\delta_{1,8,27},\delta'_{1,8})
=\Sc_{res}^m(\tau,\nu;{\rm Re}\,\delta_{1,8,27},{\rm
Re}\,\delta'_{1,8})\cdot
 \Sc_{res}^m(\tau,\nu;i\,{\rm Im}\,\delta_{1,8,27},i\,{\rm Im}\,\delta'_{1,8}).
 \label{eq:Sc_Sc}
 \en
Note that $\Sc_{res}^m(\tau,\nu;i\,{\rm Im}\,\delta_i,i\,{\rm
Im}\,\delta'_i)$ is a $n\times n$ real matrix. Substituting the
above expression of $S^{1/2}_{res}$ into Eq.~(\ref{eq:master1}),
we have
 \be
 A_{FSI}=\Sc_{res}^{1/2}(\tau,\nu;{\rm Re}\,\delta_{1,8,27},{\rm Re}\,\delta'_{1,8})\cdot
 \Sc_{res}^{1/2}(\tau,\nu;i\,{\rm Im}\,\delta_{1,8,27},i\,{\rm Im}\,\delta'_{1,8})\cdot
 A_{fac}.
 \label{eq:master2}
 \en
An overall phase in Eq.~(\ref{eq:master2}) can be removed and we
are free to set ${\rm Re}\,\delta_{27}=0$. Furthermore, in our
analysis (as well as in many analyses using naive or QCD
factorization approaches), various form factors and $m_s$ in
$A_{fac}$ are allowed to float in some given ranges of values.
Therefore, an overall scaling factor ($\exp(-{\rm
Im}\,\delta_{27})$) can be absorbed into the form factors in
$A_{fac}$ and we set ${\rm Im}\,\delta_{27}=0$ to avoid double
counting. We are left with two mixing angles, four real phase
differences and four imaginary phase differences:
 \be
 \tau,\quad
 &&\nu,
 \quad
 \delta^{(\prime)}\equiv{\rm Re}(\delta_{8^{(\prime)}}-\delta_{27}),
 \quad
 \sigma^{(\prime)}\equiv{\rm Re}(\delta_{1^{(\prime)}}-\delta_{27}),
 \nonumber
 \\
 &&\kappa^{(\prime)}\equiv{\rm Im}(\delta_{8^{(\prime)}}-\delta_{27}),
 \quad
 \xi^{(\prime)}\equiv{\rm Im}(\delta_{1^{(\prime)}}-\delta_{27}).
 \label{eq:FSIparameters}
 \en
The number of the residual FSI parameters is still quite large. It
will be preferable to reduce it through some physical arguments
or the consideration of some plausible cases.

It is interesting to see how the residual FSI behaves in a U(3)
symmetric case. It is known that the U$_A(1)$ breaking is
responsible for the mass difference between $\eta$ and $\eta'$ and U(3)
symmetry is not a good symmetry for low-lying pseudoscalars.
However, U(3) symmetry may still be a reasonable one for
a system that rescatters at energies of order $m_B$. The mass difference between $\eta$
and $\eta'$, as an indicator of U(3) symmetry breaking effect,
does not lead to sizable energy difference of these particles in
charmless $B$ decays. In the literature, some authors also make
use of U(3) symmetry in charmless $B$ decays (see, for
example~\cite{Pham:2007nt}).

The full U(3) symmetry requires:
 \be
 r_i=\bar r_i=\tilde r_i=\check r_i,
 \en
for each $i=0,a,e,t$. This imposes a major reduction of
parameters. Note that the reduction is more easier to preform in
the $r_i$ formalism than in the SU(3) decomposition formalism.
This is one of the advantages of the former formalism.

In the U(3) case, we are constrained to have (see Appendix B)
 \be
 r^{(m)}_e r^{(m)}_a=0.
 \en
Consequently, there are two different solutions:
(a)~the annihilation type ($r^{(m)}_a\neq0,\,r^{(m)}_e=0$) with
 \be
 \delta_{27}=\delta_8'=\delta'_1,
 \quad
 \delta_8,
 \quad
 \delta_1,
 \quad
 \tau=-\frac{1}{2} \sin^{-1}\frac{4\sqrt5}{9},
 \quad
 \nu=-\frac{1}{2}\sin^{-1}\frac{4\sqrt2}{9},
 \label{eq:solutionreU3ra}
 \en
and (b)~the exchange type ($r^{(m)}_e\neq0,\,r^{(m)}_a=r^{(m)}_t=0$)
with
 \be
 \delta_{27}=\delta'_8=\delta_1',
 \quad
 \delta_8=\delta_1,
 \quad
 \tau=\frac{1}{2} \sin^{-1}\frac{\sqrt5}{3},
 \quad
 \nu=\frac{1}{2}\sin^{-1}\frac{2\sqrt2}{3}.
 \label{eq:solutionreU3re}
 \en
The explicit expressions of $r^{(m)}_i$ in terms of these phases can be
found in Appendix B.

It is interesting to note that in both solutions of the U(3) case,
a common constraint
 \be
  \delta_{27}=\delta_8'=\delta'_1,
  \label{eq:constrain}
 \en
has to be satisfied. To reduce the number of the residual FSI
parameters shown in Eq.~(\ref{eq:FSIparameters}), we consider a
restricted SU(3) case, which is close, but not necessarily
identical, to the full U(3) case. Motivated by
Eq.~(\ref{eq:constrain}), we consider the parameter space around
 \be
\delta^\prime\simeq
 \sigma^\prime\simeq0,
 \qquad
 \kappa^\prime\simeq
 \xi^\prime\simeq0.
 \en
The above restriction on the FSI parameter space is a rather
strong model assumption. When comparing the fitted FSI parameters
with those in Eqs.~(\ref{eq:solutionreU3ra}) and
(\ref{eq:solutionreU3re}), it is possible to determine whether the
exchange-type, the annihilation-type or a mixed solution is
preferred by data.

\section{Numerical results}

In our numerical study, masses and lifetimes are taken from the
review of the Particle Data Group (PDG)~\cite{PDG}, and the branching ratios of $B$ to
charmless meson decays are taken
from~\cite{HFAG,Petadata,CDFnew}.
We use $f_\pi=$ 131 MeV, $f_{K}=$ 156 MeV~\cite{PDG} and
$f_{B_{(s)}}=$ 200 (230) MeV for decay constants.
The values of CKM matrix elements are taken from the central
values of the latest CKM fitter's results~\cite{CKMfitter}.

We use the QCD factorization calculated amplitudes
\cite{Beneke:2003zv} for the factorization amplitudes in the
right-hand-side of Eq.~(\ref{eq:master1}). We take the
renormalization scale $\mu=4.2$~GeV and the power correction
parameters $X_{A,H}=\ln(m_B/\Lambda_h)(1+\rho_{A,H}
e^{i\phi_{A,H}} )$. Hadronic parameters in factorization
amplitudes are fit parameters in addition to FSI parameters, and
are allowed to vary in the following ranges:
 \be
 &&\hspace{0.5cm}
 0\leq\rho_{A}=\rho_H\leq2,\quad
 -\pi<\phi_{A,H}\leq\pi,
 \quad
 m_s(2.1{\rm GeV})=(100\pm30){\rm
 MeV},\nonumber\\
 &&F^{B\pi}_0(0)=0.25\pm0.05,
 \quad
 F^{BK}_0(0)=0.35\pm0.08,
 \quad
 F^{B_sK}(0)=0.31\pm0.08.
 \label{eq:QCDFHparameters}
 \en
Note that we take $\rho_A=\rho_H$ for simplicity. These
estimations agree with those in \cite{Beneke:2003zv,LF,MS}, while
the ranges of form factors are slightly enlarged to include the
possible effect of the overall scaling factor $\exp(-{\rm
Im}\delta_{27})$ from $\Sc^{1/2}_{res}$. Other parameters (if not
specified explicitly) in the QCDF amplitudes are taken from the
central values of those used in \cite{Beneke:2003zv}. For the FSI
parameters, we set allowed ranges to be
 \be
 -\frac{\pi}{2}<\tau,\nu\leq\frac{\pi}{2},
 \quad
 -\pi<\delta,\sigma\leq\pi,
 \quad
 -0.35\leq\kappa,\xi\leq0.35
 \en
for the mixing angles, real and imaginary parts of FSI phase
differences. In the fit we take $\delta'=\sigma'=\kappa'=\xi'=0$
as mentioned in the end of the previous section. The effects of
relaxing these constraints will also be estimated.

\begin{table}[t!]
\caption{ Confidence level (C.L.), $\chi^2_{\rm min}/{\rm d.o.f.}$
and various contributions to $\chi^2_{\rm min}$ for the best
fitted solution. Numbers of data used are shown in parentheses.
 \label{tab:chisquare}
}
\begin{ruledtabular}
{\footnotesize
\begin{tabular}{cccccc}
 C. L.
 &$\chi^2_{\rm min.}/{\rm d.o.f.}$
 &$\chi^2_{\{\B(\ov B{}^0\to K\pi),\dots\}}$
 &$\chi^2_{\{\A(\ov B{}^0 \to K\pi),\dots\}}$
 &$\chi^2_{\{\B(B^-\to K\pi),\dots\}}$
 &$\chi^2_{\{\A(B^- \to K\pi),\dots\}}$
 \\
  \hline
   0.04 (43)
  & 1.51 (43)
  & 4.6 (4)
  & 1.6 (3)
  & 4.5 (4)
  & 7.0 (4)
 \\
 \hline
  $\chi^2_{\{\B(B^-\to \pi\pi),\dots\}}$
  &$\chi^2_{\{\A(B^- \to \pi\pi),\dots\}}$
  &$\chi^2_{\{B(\ov B{}^0\to \pi\pi),\dots\}}$
  &$\chi^2_{\{\A(\ov B{}^0 \to \pi\pi),\dots\}}$
  &$\chi^2_{\{\B(\ov B_s),\A(\ov B_s)\}}$
  &$\chi^2_{\{S(\ov B {}^0))\}}$
  \\
   \hline
   2.6 (4)
  & 6.8 (4)
  & 8.0 (9)
  & 2.4 (3)
  & 1.6 (4)
  & 6.0 (4)
  \end{tabular}
  }
\end{ruledtabular}
\end{table}

We perform a $\chi^2$ analysis with all available data on
CP-averaged rates and CP asymmetries in $\ov B{}_{u,d,s}\to PP$
decays. There are altogether 43 data used in the fit. The
confidence level and $\chi^2$ for the best fitted case are shown
in Table~\ref{tab:chisquare}. Contributions to $\chi^2_{\rm min.}$
from various sub-sets of data are also given. For example,
$\chi^2_{\{\B(\ov B{}^0\to K\pi),\dots\}}$ in the table denotes
the $\chi^2$ contribution obtained from 4 CP-averaged $\ov
B{}^0\to K^-\pi^+,\, \ov K {}^0\pi^0,\,\ov K{}^0\eta,\,\ov
K{}^0\eta'$ rates, which are related through FSI [see
Eq.~(\ref{eq:FSIB0Kpi}), and see Eqs.
(\ref{eq:FSIBKpi0})--(\ref{eq:FSIB0pipi}) for other groups].
Numbers of data used are shown in parentheses.

From Table~\ref{tab:chisquare}, 
%
by comparing the $\chi^2$ value and the number of data used in
each group, we are able to have a rough idea on the quality of the
fit. In most cases, the $\chi^2$ values are compatible or smaller
than the numbers of data used, indicating reasonable fit to
measurements in these groups. However, the $\chi^2_{\{\A(B^-\to
K\pi),\dots\}}$, $\chi^2_{\{\A(B^- \to \pi\pi),\dots\}}$ and
$\chi^2_{S(\ov B{}^0)}$ values are larger than the corresponding
numbers of data used. We will discuss more on the sources causing
these sizable $\chi^2$ later.

\begin{table}[t!]
\caption{ Fitted hadronic and FSI parameters. Upper table contains
fitted parameters in factorization amplitudes, while the lower one
contains fitted FSI parameters. Note that parameters with values
given in parentheses are not fitted ones (see text).
 \label{tab:parameters1}
}
\begin{ruledtabular}
\begin{tabular}{ccccccccc}
  $\rho_{A,H}$
 &$\phi_A({}^\circ)$
 &$\phi_H({}^\circ)$
 &$m_s$(MeV)
 &$F_0^{B\pi}(0)$
 &$F_0^{BK}(0)$
 &$F_0^{B_sK}(0)$
 \\
  \hline
    $1.18_{-0.23}^{+0.08}$
  & $-65.7_{-16.0}^{+16.3}$
  & $7.5_{-80.1}^{+40.6}$
  & $84.3_{-1.5}^{+1.8}$
  & $0.258_{-0.004}^{+0.017}$
  & $0.314_{-0.012}^{+0.030}$
  & $0.237_{-0.007}^{+0.025}$
  \\
  \hline
  $\tau({}^\circ)$
  &$\nu({}^\circ)$
  &$\delta({}^\circ)$
  &$\sigma({}^\circ)$
  &$\kappa$
  &$\xi$
  &$\delta'$, $\sigma'({}^\circ)$
  &$\kappa'$, $\xi'$
  \\
  \hline
    $20.6_{-1.8}^{+1.9}$
  & $41.2_{-3.8}^{+24.7}$
  & $51.4_{-26.8}^{+9.8}$
  & $88.9_{-8.9}^{+109.5}$
  & $-0.35_{-0.00}^{+0.03}$
  & $0.26_{-0.61}^{+0.09}$
  & $(0\pm10)$
  & $(0\pm0.05)$
  \\
\end{tabular}
\end{ruledtabular}
\end{table}

We give the fitted parameters in Table~\ref{tab:parameters1}.
Uncertainties are obtained by scanning the parameter space with
$\chi^2\leq\chi^2_{\rm min}+1$. The parameters consist of those in
factorization amplitude (in the upper table) and of FSI (in the
lower table). Values given in parenthesis are not fitted ones. We
take $\delta',\sigma'=(0\pm10)^\circ$ and $\kappa',\xi'=0\pm0.05$
for estimation.

We note that:
(i)~Most of our fitted values for hadronic parameters in
factorization amplitudes agree with those usually used in \cite{Beneke:2003zv,LF,MS}.
However, the fit seems to prefer a small value of
$F^{B_sK}(0)$, which is at the lower end of the allowed region
given in Eq.~(\ref{eq:QCDFHparameters}).
(ii)~Although it helps improve the fit, the effect of $\phi_H$ is
sub-leading. On the other hand, the effect of $\phi_A$ is
prominent. The fitted $\phi_A$ is around $-66^\circ$, which is
close to $-55^\circ$ as used in the so-called S4 scenario in
QCDF~\cite{Beneke:2003zv}. When turning off FSI phases, our
results should be similar to those obtained in the S4 scenario.
(iii)~ The fitted $\tau\simeq 21^\circ$ and $\nu\simeq41^\circ$
are closer to $\tau=24.1^\circ$ and $\nu=35.3^\circ$ of the
exchange-type solution~[see Eq.~(\ref{eq:solutionreU3re})] than to
$\tau=-41.8^\circ$ and $\nu=-19.5^\circ$ of the annihilation-type
solution~[see Eq.~(\ref{eq:solutionreU3ra})].
The exchange-type solution is more favorable.
(iv) In this work, residual FSI is taken as left-over FSI that
complements the FSI in factorization amplitudes. In principle,
there is a possible double counting in $\phi_{A,H}$ and residual
FSI phases. However, in practice the residual FSI is dominated by
the exchange rescattering, which provides important effects on
rates and CP asymmetries as we shall see later. These effects
cannot be easily obtained by varying $\rho_{A,H}$ and $\phi_{A,H}$
in reasonable ranges. In fact, numerically the $\chi^2$/d.o.f.
will not be reduced by freezing either of these parameters. Hence,
both parameters are numerically important.

\begin{table}[t!]
\caption{ \label{tab:table-br} Branching ratios of various
$\overline B\to PP$ modes in units of $10^{-6}$.
Fac, ``FSI" and FSI denote factorization, partial FSI and full FSI
results, respectively. See the maintext for details. Experimental
results are taken from \cite{HFAG,Petadata}.}
\begin{ruledtabular}
\begin{tabular}{lcccc}
 Mode
      &Exp
      &Fac
      &``FSI"
      &FSI
      \\
\hline
 $\ov B{}^0\to K^-\pi^+$
        & $19.4\pm0.6$
        & (16.0)
        & (22.5)
        & $20.1_{-0.3}^{+1.7}{}_{-2.5}^{+2.5}$
        \\
 $\ov B {}^0\to \ov K {}^0\pi^0$
        & $9.8\pm0.6$
        & (7.2)
        & (10.2)
        & $9.2_{-0.2}^{+0.7}{}_{-1.2}^{+1.2}$
        \\
 $\overline B {}^0\to \ov K {}^0\eta$
        & $1.0\pm0.3$
        & (0.9)
        & (1.7)
        & $1.4_{-0.1}^{+0.4}{}_{-0.4}^{+0.5}$
        \\
 $\overline B {}^0\to \ov K {}^0\eta'$
        & $64.9\pm3.1$
        & (66.4)
        & (62.3)
        & $65.9_{-10.6}^{+6.9}{}_{-8.1}^{+9.2}$
        \\
        \hline
 $B^-\to \ov K{}^0\pi^-$
        & $23.1\pm1.0$
        & (18.0)
        & (26.1)
        & $22.5_{-1.1}^{+2.6}{}_{-0.7}^{+3.0}$
        \\
 $B^-\to K^-\pi^0$
        & $ 12.9\pm 0.6 $
        & (10.1)
        & (14.3)
        & $12.4_{-0.2}^{+1.5}{}_{-1.6}^{+1.6}$
        \\
 $B^-\to K^-\eta$
        & $2.7\pm 0.3$
        & $(1.4)$
        & (2.5)
        & $2.1_{-0.1}^{+0.6}{}_{-0.5}^{+0.6}$
        \\
 $B^-\to K^-\eta'$
        & $70.2\pm 2.5$
        & (70.1)
        & (65.0)
        & $70.8_{-12.3}^{+6.6}{}_{-9.2}^{+10.3}$
        \\
        \hline
 $B^-\to \pi^-\pi^0$
        & $5.59^{+0.41}_{-0.40}$
        & $(5.18)$
        & (5.18)
        & $5.18_{-0.38}^{+0.55}{}_{-0.00}^{+0.00}$
        \\
 $B^-\to K^0 K^-$
        & $1.36^{+0.29}_{-0.27}$
        & $(1.22)$
        & (1.77)
        & $1.46_{-0.04}^{+0.35}{}_{-0.13}^{+0.15}$
        \\
 $B^-\to \pi^-\eta$
        & $4.4\pm0.4$
        & $(4.10)$
        & (4.47)
        & $4.23_{-0.23}^{+0.59}{}_{-0.37}^{+0.34}$
        \\
 $B^-\to \pi^-\eta'$
        & $2.7^{+0.6}_{-0.5}$
        & (3.09)
        & (2.76)
        & $3.31_{-0.51}^{+0.19}{}_{-0.54}^{+0.65}$
        \\
\hline
 $\ov B {}^0\to \pi^+\pi^-$
        & $5.16\pm 0.22$
        & (6.65)
        & (7.56)
        & $5.30_{-0.49}^{+1.92}{}_{-0.40}^{+0.39}$
        \\
 $\ov B {}^0\to \pi^0 \pi^0$
        & $1.55\pm0.35$\footnotemark[1]
        & (0.50)
        & (0.36)
        & $1.04_{-0.55}^{+0.12}{}_{-0.08}^{+0.10}$
        \\
 $\ov B {}^0\to \eta\eta$
        & $0.8\pm0.4(<1.4)$
        & (0.21)
        & (0.10)
        & $0.46_{-0.11}^{+0.24}{}_{-0.08}^{+0.10}$
        \\
 $\ov B {}^0\to \eta \eta'$
        & $0.5\pm0.4(<1.2)$
        & (0.22)
        & (0.24)
        & $0.88_{-0.40}^{+0.39}{}_{-0.21}^{+0.24}$
        \\
 $\ov B {}^0\to \eta'\eta'$
        & $0.9_{-0.7}^{+0.8}(<2.1)$
        & (0.16)
        & (0.30)
        & $1.06_{-0.31}^{+1.16}{}_{-0.28}^{+0.36}$
        \\
 $\ov B {}^0\to K^+ K^-$
        & $0.15^{+0.11}_{-0.10}$
        & (0.09)
        & (0.05)
        & $0.10_{-0.02}^{+0.35}{}_{-0.06}^{+0.10}$
        \\
 $\ov B {}^0\to K^0\ov K^0$
        & $0.96^{+0.21}_{-0.19}$
        & (1.47)
        & (1.56)
        & $1.10_{-0.12}^{+0.46}{}_{-0.11}^{+0.12}$
        \\
 $\ov B {}^0\to \pi^0 \eta$
        & $0.9\pm0.4(<1.5)$
        & $(0.26)$
        & (0.37)
        & $0.31_{-0.01}^{+0.05}{}_{-0.06}^{+0.06}$
        \\
 $\ov B {}^0\to \pi^0\eta'$
        & $1.2\pm0.7$\footnotemark[2]
        & (0.32)
        & (0.22)
        & $0.42_{-0.15}^{+0.02}{}_{-0.11}^{+0.13}$
        \\
 \end{tabular}
 \footnotetext[1]{An $S$ factor of 1.8 is included in the uncertainty.}
 \footnotetext[2]{An $S$ factor of 1.7 is included in the uncertainty.}
\end{ruledtabular}
\end{table}

\subsection{Rates in $\overline B {}^0$
and $B^-$ Decays}

In Table~\ref{tab:table-br}, we show the CP-average rates of $\ov
B{}^0,B^-\to PP$ decays. In the table, Fac, ``FSI" and FSI denote
factorization, partial FSI and full FSI results, respectively. The
FSI results are obtained with the best-fitted parameters shown in
Table~\ref{tab:parameters1}. The factorization results are
obtained by using the same set of the best fitted parameters, but
with the residual FSI phases ($\delta^{(\prime)}$,
$\sigma^{(\prime)}$, $\kappa^{(\prime)}$ and $\xi^{(\prime)}$) set
to zero, while the partially FSI results are obtained similarly,
but only with the real FSI phases ($\delta^{(\prime)}$,
$\sigma^{(\prime)}$) set to zero. Recall that in
Eq.~(\ref{eq:master2}) the $\Sc^{1/2}_{res}$ matrix can be
factorized into two parts, one involving real FSI phases and the
other involving imaginary phases. The ``FSI" results only make use
of the one involving imaginary phases and are sort of ``half-way"
from the factorization results to the full FSI ones.

Uncertainties for factorization results are not given and can be
found elsewhere (for example, in \cite{Beneke:2003zv}). The first
uncertainties in FSI results are obtained by scanning the
parameter space with $\chi^2\leq\chi^2_{\rm min}+1$, while keeping
$\delta'=\sigma'=0$ and $\kappa'=\xi'=0$. The second uncertainties
in FSI results are from the variations of $\delta^{\prime}$,
$\sigma^{\prime}$, $\kappa^{\prime}$ and $\xi^{\prime}$.
From the table, we see that the $\ov B\to \ov K\pi$ and $\ov B\to
\ov K\eta^{(\prime)}$ rates are quite sensitive to
$\delta^{\prime}$, $\sigma^{\prime}$, $\kappa^{\prime}$ and
$\xi^{\prime}$. Hence, a larger variation of these parameters is
not preferred by the data.

As shown in Table~\ref{tab:table-br}, the residual FSI results
agree with data. Before turning on the residual FSI, the
factorization results are close to the S4 ones as expected. After
the residual FSI is turned on, some rates are enhanced remarkably. In
particular, $\ov B{}^0$ decays in the $\Delta S=0$ transitions
receive large contributions from the residual FSI. In the
following, we will focus on effects of the FSI on some interesting
modes.

\begin{figure}[t!]
\centerline{\DESepsf(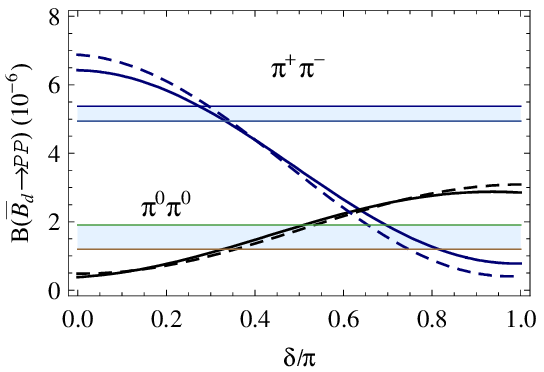 width
8cm)\hspace{1cm}\DESepsf(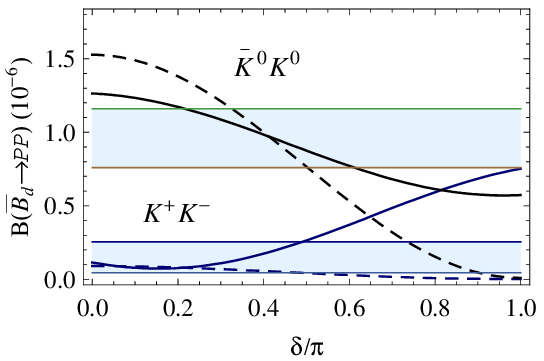 width 8.2cm)}
\caption{$\ov B {}^0\to \pi^+\pi^-$, $\pi^0\pi^0$ rates (left)
and $\ov B {}^0\to K^+K^-$, $\ov K {}^0 K^0$ rates (right) versus
$\delta$ are plotted. The solid line is obtained by using all other
parameters set to their best-fitted values, while the dashed line is
obtained using the exchange-type solution for FSI parameters (see
text). Bands are one-sigma ranges of experimental data.
Theoretical uncertainties are not shown. Note that the fitted
$\delta/\pi$ is around $0.3$ (see Table~\ref{tab:parameters1}).}
\label{fig:B0Br}
\end{figure}

Through the residual FSI, $\ov B{}^0\to\pi^+\pi^-$ and
$\pi^0\pi^0$ rates~\footnote{For the factorization amplitudes, we
use the central values of Gegenbauer coefficients for the pion
wave function, $\alpha^\pi_2(2\,{\rm GeV})=0.2\pm0.1$, used in
\cite{Gegenbauer} and do not consider the case of using a larger
Gegenbauer coefficient, which leads to a larger $\pi^0\pi^0$
rate.} are reduced and enhanced roughly by factor $2$,
respectively, leading to a better agreement with data. Note that
in the ``FSI" case, the $\pi^+\pi^-$ rate is enhanced, while the
$\pi^0\pi^0$ rate is slightly reduced. Both of them are pushed
even further from the data. There are the real FSI phases
($\delta,\sigma$) that will change these rates in the right
direction.

In Fig.~\ref{fig:B0Br}, we show the $\ov B{}^0\to\pi^+\pi^-$ and
$\pi^0\pi^0$ rates versus $\delta$. The solid line is obtained by
using all other parameters set to their best-fitted values, while
the dashed line is obtained using the exchange-type solution for
FSI parameters [see, Eq.~(\ref{eq:solutionreU3re})] with $\tau$,
$\nu$ fixed, $\sigma=\delta$ and $\kappa=\xi$ taken from the
average of the central values of the fitted $\kappa$ and $\xi$. We
see that $\ov B{}^0\to\pi^+\pi^-$ and $\pi^0\pi^0$ rates are
reduced and enhanced, respectively, as $\delta$ is increasing.
Both rates reach the measured ones at $\delta\sim0.3\pi$.

\begin{figure}[t!]
 \centerline{\DESepsf(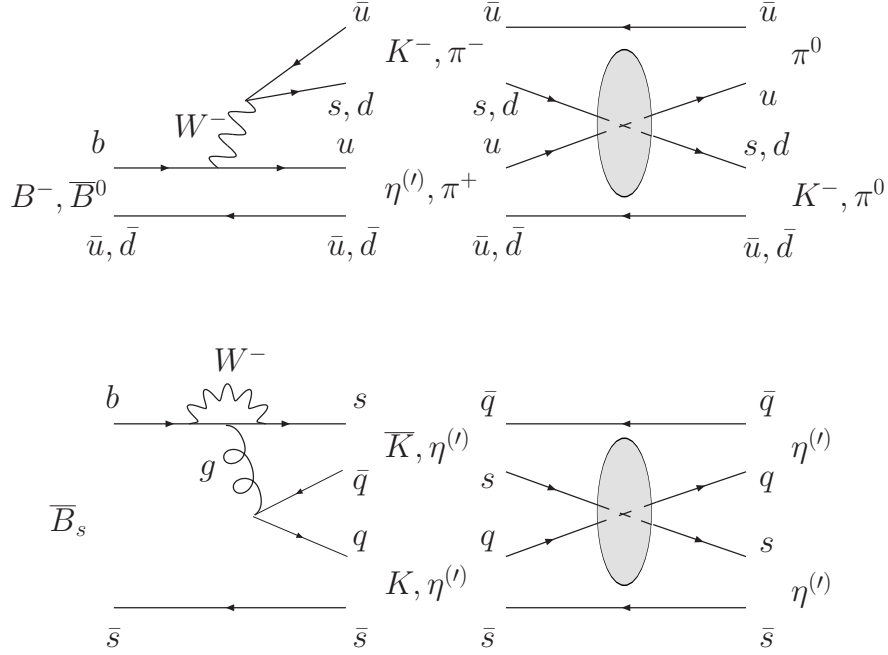 width 11cm)}
\caption{Exchange rescattering in $\ov B{}^0\to\pi^0\pi^0$,
$B^-\to K^-\pi^0$ and $\ov B_s\to\eta^{(\prime)}\eta^{(\prime)}$
decays.} \label{fig:re}
\end{figure}

It is known that in order to have the $\pi^0\pi^0$ rate as large as
observed, we need a sizable color-suppressed tree
amplitude~\cite{largeC}. In the residual FSI, a large
color-suppressed tree contribution can be generated from the exchange
rescattering. As shown in the upper part of Fig.~\ref{fig:re}, the
color-allowed tree amplitude of the $\ov B{}^0\to\pi^+\pi^-$
decay, a main FSI source in this sector, can produce a
color-suppressed tree amplitude for the $\ov B{}^0\to\pi^0\pi^0$
decay through the exchange rescattering. At the same time, the
$\pi^+\pi^-$ rate is reduced as it rescatters. We see that the
exchange rescattering is responsible for the enhancement of $\pi^0\pi^0$ and the suppression of
$\pi^+\pi^-$.

In Fig.~\ref{fig:B0Br}, we show the $\ov B{}^0\to K^+ K^-$ and
$\ov K{}^0 K^0$ rates. It is known that the $K^+K^-$ rate is
sensitive to annihilation-type rescattering~\cite{quasielastic}
(corresponding to the $r_a$ and $r_t$ terms as depicted in
Fig~\ref{fig:r}(c) and (d)). In the SU(3) case (solid line), for
$\delta\leq \pi/2$, the $K^+K^-$ constraint can be easily
satisfied, while in the U(3) case (dashed line), the constraint on
$\delta$ is even weaker. These features are understandable, since
in both cases, the exchange-type rescattering, which cannot
generate $K^+K^-$ final state by rescattering the $\ov
B{}^0\to\pi^+\pi^-$ decay amplitude, is dominating. Note that the
$\ov K{}^0K^0$ rate is reduced through FSI, giving better
agreement with data without violating the $K^+K^-$ bound.

In summary, the residual FSI improves the agreement between
theory and experiment for rates, in particularly, it
resolves the discrepancy between data and theoretical
expectations on $\ov B{}^0\to\pi^+\pi^-$ and $\pi^0\pi^0$ rates.

\begin{table}[t!]
\caption{ \label{tab:table-acp} Same as Table~\ref{tab:table-br},
except for the direct CP asymmetries $\A$ (in units of percent)
in various $\overline B\to PP$ modes. }
\begin{ruledtabular}
\begin{tabular}{lcrrr}
 Mode
      &Exp
      &Fac
      &``FSI"
      &FSI
      \\
\hline
 $\ov B{}^0\to K^-\pi^+$
        & $-9.8^{+1.2}_{-1.1}$
        & $(-11.8)$
        & $(-12.2)$
        & $-9.0_{-0.6}^{+2.0}{}_{-2.2}^{+2.0}$
        \\
 $\ov B {}^0\to \ov K {}^0\pi^0$
        & $-1\pm13$\footnotemark[1] 
        & $(3.3)$
        & $(0.9)$
        & $-12.8_{-1.0}^{+2.2}{}_{-1.5}^{+1.7}$
        \\
 $\overline B {}^0\to \ov K {}^0\eta$
        & --
        & (10.7)
        & $(2.1)$
        & $-28.7_{-1.9}^{+8.0}{}_{-1.9}^{+3.3}$
        \\
 $\overline B {}^0\to \ov K {}^0\eta'$
        & $4.8\pm5.1$
        & ($0.2$)
        & $(0.6)$
        & $1.7_{-0.2}^{+0.8}{}_{-0.4}^{+0.3}$
        \\
        \hline
 $B^-\to \ov K{}^0\pi^-$
        & $0.9\pm2.5$
        & (0.3)
        & $(0.2)$
        & $-0.3_{-0.6}^{+0.7}{}_{-1.1}^{+1.2}$
        \\
 $B^-\to K^-\pi^0$
        & $5.0\pm 2.5 $
        & $(-11.8)$
        & $(-10.3)$
        & $4.8_{-1.2}^{+1.4}{}_{-2.0}^{+1.9}$
        \\
 $B^-\to K^-\eta$
        & $-27\pm 9$
        & (39.8)
        & (28.2)
        & $-27.3_{-3.0}^{+8.6}{}_{-6.3}^{+10.8}$
        \\
 $B^-\to K^-\eta'$
        & $1.6\pm 1.9 $
        & $(-2.6)$
        & $(-2.6)$
        & $-3.3_{-0.5}^{+1.0}{}_{-0.5}^{+0.5}$
        \\
        \hline
 $B^-\to \pi^-\pi^0$
        & $6\pm5$
        & $(-0.06)$
        & $(-0.06)$
        & $-0.06_{-0.01}^{+0.00}{}_{-0.00}^{+0.00}$
        \\
 $B^-\to K^0 K^-$
        & $12^{+17}_{-18}$
        & $(-3.5)$
        & $(-1.8)$
        & $12.8_{-12.8}^{+9.1}{}_{-17.8}^{+16.0}$
        \\
 $B^-\to \pi^-\eta$
        & $-16\pm7$
        & (19.7)
        & $(22.0)$
        & $-12.3_{-2.9}^{+4.1}{}_{-3.2}^{+3.5}$
        \\
 $B^-\to \pi^-\eta'$
        & $21\pm15$
        & (22.8)
        & (20.3)
        & $54.8_{-10.6}^{+5.3}{}_{-3.0}^{+1.7}$
        \\
\hline
 $\ov B {}^0\to \pi^+\pi^-$
        & $38\pm15$\footnotemark[2] 
        & (22.3)
        & (21.1)
        & $15.5_{-4.3}^{+10.2}{}_{-4.5}^{+4.6}$
        \\
 $\ov B {}^0\to \pi^0 \pi^0$
        & $43^{+25}_{-24}$
        & $(-51.5)$
        & $(-45.8)$
        & $48.3_{-33.1}^{+11.5}{}_{-13.1}^{+11.8}$
        \\
 $\ov B {}^0\to \eta\eta$
        & --
        & $(-11.7)$
        & $(-77.6)$
        & $-50.7_{-12.4}^{+15.0}{}_{-16.3}^{+15.7}$
        \\
 $\ov B {}^0\to \eta \eta'$
        & --
        & $(-28.5)$
        & $(-29.3)$
        & $-5.7_{-22.2}^{+9.5}{}_{-7.4}^{+7.8}$
        \\
 $\ov B {}^0\to \eta'\eta'$
        & --
        & (3.6)
        & $(18.7)$
        & $29.7_{-1.7}^{+26.2}{}_{-6.6}^{+8.3}$
        \\
 $\ov B {}^0\to K^+ K^-$
        & --
        & (0)
        & (52.9)
        & $71.0_{-41.4}^{+10.9}{}_{-15.6}^{+20.6}$
        \\
 $\ov B {}^0\to K^0\ov K^0$
        & $-58^{+73}_{-66}$
        & $(-9.0)$
        & $(-19.9)$
        & $-37.8_{-37.1}^{+\,\,\,8.4}{}_{-15.0}^{+15.2}$
        \\
 $\ov B {}^0\to \pi^0 \eta$
        & --
        & (19.7)
        & $(19.1)$
        & $7.2_{-13.8}^{+11.5}{}_{-0.5}^{+0.4}$
        \\
 $\ov B {}^0\to \pi^0\eta'$
        & --
        & (13.2)
        & (12.4)
        & $22.7_{-20.5}^{+\,\,\,7.7}{}_{-1.0}^{+1.0}$
        \\
 \end{tabular}
  \footnotetext[1]{An $S$ factor of 1.4 is included in the uncertainty.}
  \footnotetext[2]{An $S$ factor of 2.4 is included in the uncertainty.}
\end{ruledtabular}
\end{table}

\subsection{Direct CP Violations in $\overline B {}^0$
and $B^-$ Decays}

Results for direct CP asymmetries in $\ov B{}^0,B^-\to PP$
decays are summarized in Table~\ref{tab:table-acp}. In general, the residual FSI
has a large impact on direct CP violations of many modes. In the
following, we will focus on some interesting results.

We first concentrate on the modes that lead to the $K\pi$ puzzle
in direct CP violation. We see that before the residual FSI is
turned on (i.e. taking $\Sc_{res}=1$), we have $\A(\ov B{}^0\to
K^-\pi^+)\simeq\A(B^-\to K^-\pi^0)\simeq-0.12$ from the
annihilation amplitude for $\phi_A(\simeq-66^\circ)$. After
turning on the residual FSI $(\Sc_{res}\neq 1)$, the asymmetry
$\A(\ov B{}^0\to K^-\pi^+)$ changes from $\sim -0.12$ to $\sim
-0.09$, while $\A(B^-\to K^-\pi^0)$ changes from $\sim -0.12$ to
$\sim +0.05$, reproducing the experimental results. In other
words, the residual FSI modifies $\A(\ov B{}^0\to K^-\pi^+)$ and
$\A(B^-\to K^-\pi^0)$ by an amount of $\sim+0.03$ and $\sim+0.17$,
respectively.

\begin{figure}[t!]
\centerline{\DESepsf(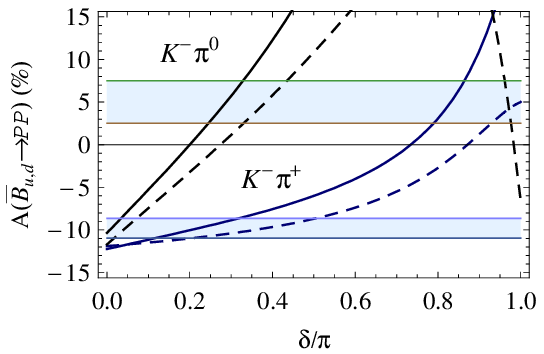 width
8cm)\hspace{1cm}\DESepsf(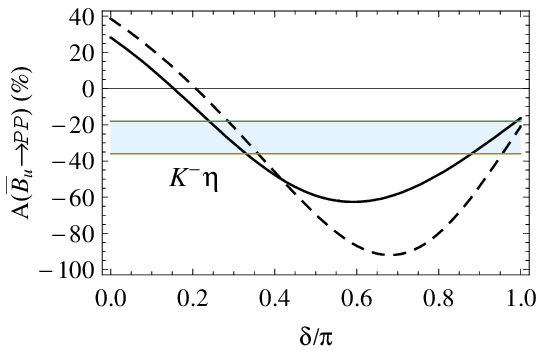 width 8cm)}
\caption{Same as Fig.~\ref{fig:B0Br}, except that direct CP
violations of $\ov B {}^0\to K^-\pi^+$ and $B^-\to K^-\pi^0$
(left), and $\ov B {}^0\to K^-\eta$ (right) versus $\delta$ are
plotted. Note that the best fitted $\delta/\pi$ is around $0.3$.}
\label{fig:Kpiacp}
\end{figure}

The residual FSI has a more prominent effect on $\A(B^-\to
K^-\pi^0)$, and, hence, it is capable of lifting the degeneracy of
$\A(B^-\to K^-\pi^0)$ and $\A(\ov B{}^0\to K^-\pi^+)$.
As shown in Fig.~\ref{fig:Kpiacp}, it only takes a small amount of $\delta$
($\sim0.2\pi$) to flip the sign of $\A(B^-\to K^-\pi^0)$, but a
large $\delta$ ($\gtrsim0.8\pi$) would be needed to do the same
thing on $\A(\ov B{}^0\to K^-\pi^+)$.

It is known that a sizable and complex color-suppressed tree
amplitude in the $B^-\to K^-\pi^0$ decay can solve the $K\pi$
puzzle~\cite{largeC}. As depicted in Fig.~\ref{fig:re}, a
color-suppressed tree amplitude in the $K^-\pi^0$ mode can be
generated from the exchange rescattering of $B^-\to
K^-\eta^{(\prime)}$ color-allowed tree amplitudes, which are known
to be quite sizable~\cite{Beneke:2003zv}. The rescattering leads
to the desired large and complex color-suppressed amplitude in the
$K^-\pi^0$ mode.

We note that in order to solve the $K\pi$ direct CP puzzle, both
$\phi_A$ and $\delta$ phases are needed. For example, a similar
analysis using rescattering among naive factorization amplitudes
that lack a large annihilation strong phase, was unable to
remove the degeneracy of  $\A(K^-\pi^+)$ and $\A(K^-\pi^0)$~\cite{quasielastic}.
In other words, rescattering from both in-elastic channels and
$PP$ final states contribute to $\A(\ov K\pi)$s, reproducing the
experimental results and resolving the $K\pi$ direct CP violation
puzzle without the need of introducing any new physics
contribution.

As noted in the previous section, the exchange rescattering is
also responsible for the enhancement of the $\ov B{}^0\to
\pi^0\pi^0$ rate. In Fig.~\ref{fig:correlation}, we show a
two-dimensional plot, exhibiting the correlation of the ratio
$\B(\ov B{}^0\to\pi^0\pi^0)/\B(\ov B{}^0\to\pi^+\pi^-)$ with the
difference $\Delta\A\equiv\A(\ov B{}^0\to K^-\pi^+)-\A(B^-\to
K^-\pi^0)$. The light shielded area is obtained by scanning over
$-\pi<\delta,\sigma\leq\pi$ and one-sigma ranges of $\tau$, $\nu$,
$\kappa$ and $\xi$, while keeping all other hadronic parameters at
their best-fitted values. The dark shielded area corresponds to
the exchange-type U(3) case and is obtained by scanning over
$-\pi<\delta=\sigma\leq\pi$, $-0.35\leq\kappa=\xi\leq0.35$, while
using $\tau$ and $\nu$ given in Eq.~(\ref{eq:solutionreU3re}) and
keeping all other hadronic parameters at their best-fitted values.
The solid line is obtained in a similar manner except keeping
$\kappa=\xi=-0.05$, which is the average of the central values of
the best fitted $\kappa$ and $\xi$. Note that in this case, only
one FSI parameter $\delta$ is varied. From the plot, we clearly
see that the data can be easily reproduced and the exchange
rescattering is responsible for generating sizable and complex
color-suppressed tree amplitudes that account for the difference
$\Delta\A$ and the $\B(\ov B{}^0\to\pi^0\pi^0)/\B(\ov
B{}^0\to\pi^+\pi^-)$ ratio at the same time.

\begin{figure}[t!]
\centerline{\DESepsf(correlation width 9cm)}
\caption{Correlation of the ratio $\B(\ov
B{}^0\to\pi^0\pi^0)/\B(\ov B{}^0\to\pi^+\pi^-)$ with the
difference $\Delta\A\equiv\A(K^-\pi^+)-\A(K^-\pi^0)$. The light
shielded area corresponds to the restricted SU(3) case, the dark
shielded area corresponds to the exchange-type U(3) case and the
solid line is the same as the previous one except keeping
$\kappa=\xi=-0.05$. See the main text for more details.}
\label{fig:correlation}
\end{figure}

\begin{table}
\caption{ \label{tab:table-Bs} Same as Table~\ref{tab:table-br},
except for the branching ratios (upper table) in the unit of
$10^{-6}$ and direct CP asymmetries (lower table) in the unit of
percent for various $\overline B_s\to PP$ modes. Experimental
results are from \cite{HFAG,CDFnew}.}
\begin{ruledtabular}
\begin{tabular}{lcccr}
 Mode
      &Exp
      &Fac
      &``FSI"
      &FSI\,\,\,\,\,\,\,
      \\
\hline
 $\B(\ov B_s{}^0\to K^-\pi^+)$
        & $5.00\pm1.25$
        & (4.72)
        & (6.08)
        & $4.81_{-0.39}^{+1.57}{}_{-0.22}^{+0.20}$
        \\
 $\B(\ov B_s {}^0\to \ov K {}^0\pi^0)$
        & --
        & (0.68)
        & (0.59)
        & $1.13_{-0.33}^{+0.24}{}_{-0.04}^{+0.05}$
        \\
 $\B(\ov B_s {}^0\to \ov K {}^0\eta)$
        & --
        & $(0.28)$
        & $(0.21)$
        & $0.59_{-0.16}^{+0.10}{}_{-0.04}^{+0.04}$
        \\
 $\B(\ov B_s {}^0\to \ov K {}^0\eta')$
        & --
        & $(2.33)$
        & (2.11)
        & $2.44_{-0.44}^{+0.14}{}_{-0.36}^{+0.42}$
        \\
        \hline
 $\B(\ov B_s {}^0\to \pi^+\pi^-)$
        & $0.53\pm 0.51$
        & $(0.30)$
        & $(0.10)$
        & $0.86_{-0.19}^{+1.72}{}_{-0.85}^{+2.93}$
        \\
 $\B(\ov B_s {}^0\to \pi^0 \pi^0)$
        & --
        & $(0.15)$
        & $(0.05)$
        & $0.43_{-0.10}^{+0.86}{}_{-0.43}^{+1.47}$
        \\
 $\B(\ov B_s {}^0\to \eta\eta)$
        & --
        & $(17.5)$
        & $(21.3)$
        & $20.2_{-1.2}^{+7.6}{}_{-4.5}^{+5.9}$
        \\
 $\B(\ov B_s {}^0\to \eta \eta')$
        & --
        & $(70.8)$
        & (65.7)
        & $63.6_{-9.2}^{+47.1}{}_{-9.7}^{+13.7}$
        \\
 $\B(\ov B_s {}^0\to \eta'\eta')$
        & --
        & $(81.9)$
        & (85.3)
        & $99.1_{-72.3}^{+6.9}{}_{-13.4}^{+15.2}$
        \\
 $\B(\ov B_s {}^0\to K^+ K^-)$
        & $24.4\pm4.8$
        & $(24.7)$
        & (25.3)
        & $20.7_{-2.1}^{+11.5}{}_{-3.0}^{+3.3}$
        \\
 $\B(\ov B_s {}^0\to K^0\ov K^0)$
        & --
        & (25.4)
        & (27.1)
        & $20.4_{-1.8}^{+12.1}{}_{-3.4}^{+3.8}$
        \\
 $\B(\ov B_s {}^0\to \pi^0 \eta)$
        & --
        & (0.06)
        & $(0.09)$
        & $0.09_{-0.00}^{+0.03}{}_{-0.00}^{+0.00}$
        \\
 $\B(\ov B_s {}^0\to \pi^0\eta')$
        & --
        & (0.09)
        & (0.11)
        & $0.13_{-0.00}^{+0.03}{}_{-0.01}^{+0.01}$
        \\
        \hline
        \hline
 $\A(\ov B_s{}^0\to K^-\pi^+)$
        & $39\pm17$
        & (33.4)
        & (36.7)
        & $26.6_{-5.2}^{+2.7}{}_{-4.7}^{+4.8}$
        \\
 $\A(\ov B_s {}^0\to \ov K {}^0\pi^0)$
        & --
        & $(-49.1)$
        & $(-46.8)$
        & $45.5_{-12.6}^{+30.7}{}_{-10.5}^{+10.1}$
        \\
 $\A(\ov B_s {}^0\to \ov K {}^0\eta)$
        & --
        & $(2.0)$
        & $(-3.5)$
        & $76.4_{-5.1}^{+14.9}{}_{-7.7}^{+6.0}$
        \\
 $\A(\ov B_s {}^0\to \ov K {}^0\eta')$
        & --
        & $(2.5)$
        & $(-2.9)$
        & $-14.6_{-21.8}^{+4.3}{}_{-4.2}^{+5.7}$
        \\
        \hline
 $\A(\ov B_s {}^0\to \pi^+\pi^-)$
        & --
        & $(0)$
        & (-22.7)
        & $-6.1_{-1.2}^{+9.7}{}_{-21.5}^{+56.4}$
        \\
 $\A(\ov B_s {}^0\to \pi^0 \pi^0)$
       & --
        & $(0)$
        & (-22.7)
        & $-6.1_{-1.2}^{+9.7}{}_{-21.5}^{+56.4}$
        \\
 $\A(\ov B_s {}^0\to \eta\eta)$
        & --
        & $(1.6)$
        & (1.1)
        & $-3.6_{-1.6}^{+2.6}{}_{-1.4}^{+1.9}$
        \\
 $\A(\ov B_s {}^0\to \eta \eta')$
        & --
        & $(0.4)$
        & (0.5)
        & $0.2_{-0.1}^{+1.7}{}_{-1.0}^{+1.1}$
        \\
 $\A(\ov B_s {}^0\to \eta'\eta')$
        & --
        & $(0.2)$
        & $(0.0)$
        & $0.0_{-3.5}^{+0.2}{}_{-0.3}^{+0.4}$
        \\
 $\A(\ov B_s {}^0\to K^+ K^-)$
        & --
        & $(-11.9)$
        & $(-12.7)$
        & $-11.0_{-1.3}^{+3.1}{}_{-2.9}^{+2.7}$
        \\
 $\A(\ov B_s {}^0\to K^0\ov K^0)$
        & --
        & (0.3)
        & $(1.1)$
        & $2.2_{-0.3}^{+1.8}{}_{-1.1}^{+1.2}$
        \\
 $\A(\ov B_s {}^0\to \pi^0 \eta)$
        & --
        & $(3.9)$
        & $(4.7)$
        & $82.8_{-20.0}^{+5.5}{}_{-4.9}^{+4.2}$
        \\
 $\A(\ov B_s {}^0\to \pi^0\eta')$
        & --
        & (37.5)
        & (33.5)
        & $93.9_{-15.5}^{+2.7}{}_{-4.4}^{+3.2}$
        \\
 \end{tabular}
\end{ruledtabular}
\end{table}

We now continue to discuss FSI effects on direct CP
asymmetries. There are several interesting results and remarks: (i)
Large effects of residual FSI on $\A$ for several other modes are
obtained. Direct CP asymmetries in $\ov B {}^0\to \ov K {}^0\pi^0$,
$\ov K{}^0\eta$, $\pi^0\pi^0$ decays and in $B^-\to K^-\pi^0$,
$K^-\eta$, $\pi^-\eta$ decays change signs in the presence of FSI.
In Fig.~\ref{fig:Kpiacp}, we see that $\A(B^-\to K^-\eta)$ is
quite sensitive to FSI. The solid line passes through the
one-sigma range of data around $\delta\sim0.3\pi$.
(ii) Recall that in Table~\ref{tab:chisquare}, we have
$\chi^2_{\{\A(B^-\to K\pi),\dots\}}=7.0$ and $\chi^2_{\{\A(B^-\to
\pi\pi),\dots\}}=6.8$ from $\A(B^-\to\ov
K{}^0\pi^-,K^-\pi^0,K^-\eta,K^-\eta')$ and
$\A(B^-\to\pi^-\pi^0,K^0K^-,\pi^-\eta,\pi^-\eta')$ results,
respectively. We see from Table~\ref{tab:table-acp} that the main
contributions to these $\chi^2$ are from $\A(B^-\to K^-\eta')$ and
$\A(B^-\to\pi^-\eta')$, respectively.
(iii) Note that the experimental uncertainty of $\A(\ov B{}^0\to
\pi^+\pi^-)$ is enlarged by a PDG $S$-factor originated from two
different measurements: $0.25\pm0.08\pm0.02$ and
$0.55\pm0.08\pm0.05$ from BaBar~\cite{Petadata} and
Belle~\cite{Ishino:2006if}, respectively. Our fitted result of
$\A(\ov
B{}^0\to\pi^+\pi^-)=(15.5_{-4.3}^{+10.2}{}_{-4.5}^{+4.6})\%$
prefers the BaBar data.
(iv) The direct CP violation of $\ov B{}^0\to\pi^0\pi^0$ flips
sign, resulting a large and positive $\A(\pi^0\pi^0)$.
(v) The direct CP violation of $B^-\to \pi^-\pi^0$ is very small
and does not receive any contribution from the residual
rescattering, since it can only rescatter into itself [see
Eq.~(\ref{eq:FSIBpipi0})]. The smallness of $\A(B^-\to
\pi^-\pi^0)$ is consistent with a requirement followed from the
CPT theorem~\cite{CPT}. The $\A(B^-\to\pi^-\pi^0)$ measurement
remains as a clean way to search for new physics
effects~\cite{Cheng:2004ru}.

\begin{figure}[t!]
\centerline{\DESepsf(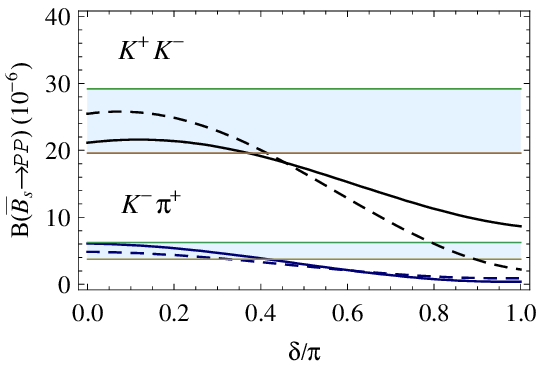 width
8cm)\hspace{1cm}\DESepsf(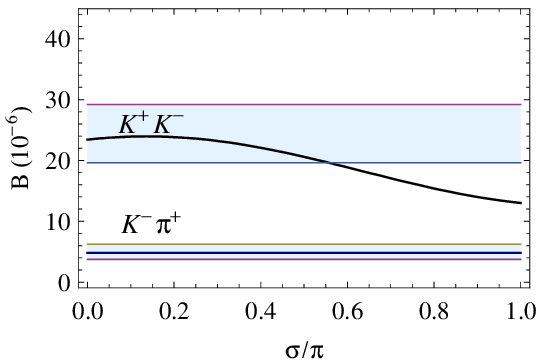 width 8cm)}
\centerline{\DESepsf(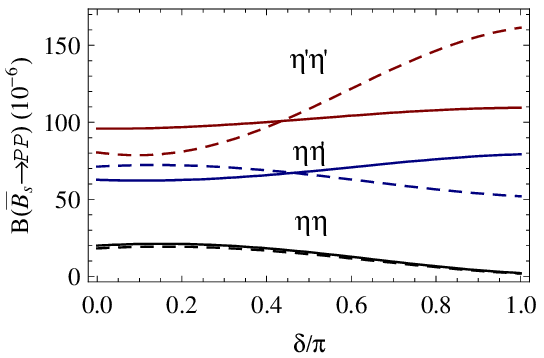 width
8cm)\hspace{1cm}\DESepsf(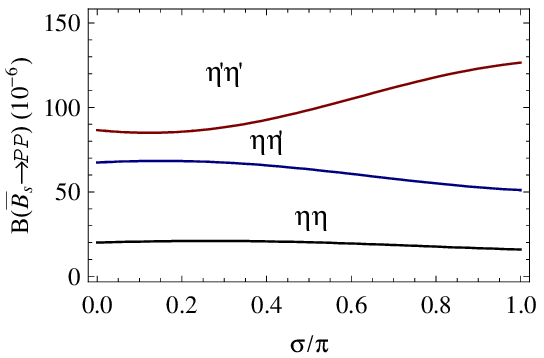 width 8cm)}
 \vspace{-0.3cm}
\caption{Same as Fig.~\ref{fig:B0Br} except that $\ov B_s{}^0\to
K^-\pi^+$ and $K^+K^-$ rates (top), $\ov
B_s{}^0\to\eta^{(\prime)}\eta^{(\prime)}$ rates (bottom) versus
$\delta$ (left) or $\sigma$ (right) (with all other parameters
fixed at the best-fitted values) are plotted here. Theoretical
uncertainties are not shown. Note that the best fitted values for
these FSI phases are $\delta/\pi\sim0.3$ and $\sigma/\pi\sim
0.5$.} \label{fig:Bs}
\end{figure}

\subsection{Rates and Direct CP asymmetries in $\ov B{}^0_s$
Decays}

We now turn to $B_s$ decays. In Table~\ref{tab:table-Bs}, we show
the CP-averaged rates and direct CP violations of $\ov B{}^0_s\to
PP$ decays. The results are then compared with data. From the table, we
see that:
(i) The $\B(\ov B_s\to K^-\pi^+)$ rate agrees well with data. From
Fig.~\ref{fig:Bs}, we note that the result is in agreement with
data for $0<\delta<\pi/2$ and any value of $\sigma$.
(ii) The $\B(\ov B_s\to K^+K^-)$ rate plotted in Fig.~\ref{fig:Bs}
versus $\delta$ and $\sigma$ agrees with data.
(iii) The $\B(\ov B_s\to \pi^+\pi^-)$ data can be reproduced, but
the result has a large uncertainty.
(iv) The $\A(\ov B_s\to K^-\pi^+)$ data can be reproduced, but the
fitted value is close to the lower end of the data.

We expect the residual FSI to have sizable contributions to
various $\ov B_s\to PP$ decay rates. For example, from
Fig.~\ref{fig:re} we see that the $\ov
B_s\to\eta^{(\prime)}\eta^{(\prime)}$ decays also receive
contributions from the exchange rescattering. Plots of $\ov B_s\to
\eta^{(\prime)}\eta^{(\prime)}$ rates versus $\delta$ and $\sigma$
are shown in Fig.~\ref{fig:Bs}. The $\ov B_s\to \eta'\eta'$ rate
is quite sensitive to the FSI phase $\sigma$. As shown in
Table~\ref{tab:table-Bs}, the $\ov B{}^0_s\to\eta'\eta'$ branching
ratio is enhanced by a factor of $1.2$ and reaches $1.0\times
10^{-4}$, which can be checked in the near future. The $\ov B_s\to
\ov K{}^0\pi^0$ and $\ov K{}^0\eta$ modes are also quite sensitive
to the residual rescattering and their rates are enhanced by
factors of 1.5 to 2, respectively.

Similar to $\ov B_{u,d}$ cases, the residual FSI also has large
impacts on many $\A(\ov B_s\to PP)$. From
Table~\ref{tab:table-Bs}, we see that signs of $\A(\ov B_s\to \ov
K{}^0\pi^0)$ and $\A(\ov B_s\to \ov K{}^0\eta^\prime)$ are
flipped.
Note that direct CP asymmetries in $\ov B_s\to \ov K{}^0\pi^0$,
$\ov K{}^0\eta$, $\pi^0\eta$ and $\pi^0\eta'$ decays are close to
or greater than 50\%. On the contrary, direct CP asymmetries in
penguin dominated $b\to s$ transition modes, such as $\ov B_s\to
K^0\ov K{}^0$, $\eta\eta'$ and $\eta'\eta'$ decays, are predicted
to be quite small.
It should be noted that these results may be subject to some
small SU(3) breaking effects.

There are increasing interests in the $\ov B_s$
sector~\cite{UTfit,PDG}. It is expected that more data from CDF
and other detectors should be available soon. Predictions on $\ov B_s$
decay rates and direct CP violations given here can be tested in
the near future.

\subsection{Time-dependent CP violations in $\ov B{}^0$ and $\ov
B{}^0_s$ Decays}

Results on time-dependent CP-asymmetries $S$ are given in
Table~\ref{tab:table-S}. The sources of the first two
uncertainties are the same as those in previous tables, while the
last uncertainty comes from the variation of
$\gamma/\phi_3=(67.6^{+2.8}_{-4.5})^\circ$~\cite{CKMfitter}.
We fit to data on mixing induced CP asymmetries. Note that for the
$\ov B{}^0\to K^0\ov K{}^0$ mode, the mixing induced CP asymmetry
obtained by BaBar
($-1.28^{+0.80+0.11}_{-0.73-0.16}$~\cite{Aubert:2006gm}) and Belle
($-0.38^{+0.69}_{-0.77}\pm0.08$~\cite{Abe:2007xd}) are quite
different, and the central value of the former exceeds the physical
range. Consequently, for this mode, only the Belle result is used
in our fit.

\begin{table}
\caption{ \label{tab:table-S} Results on the time-dependent CP
asymmetry $S$ of various $\overline B_{d,s}\to PP$ modes. The
first two uncertainties are same as those in previous tables,
while the last uncertainty comes from the variation of
$\gamma/\phi_3$. }
\begin{ruledtabular}
\begin{tabular}{lcrrr}
 Mode
      &Exp
      &Fac~$\quad$
      &``FSI"~$\,\,$
      &FSI~$\qquad$
      \\
\hline
 $\ov B {}^0\to K_S\pi^0$
        & $0.58\pm0.17$
        & $(0.780)$
        & $(0.747)$
        & $0.778_{-0.037}^{+0.003}{}_{-0.013}^{+0.014}{}_{-0.002}^{+0.003}$
        \\
 $\ov B{}^0\to K_S\eta$
        & --
        & $(0.831)$
        & $(0.772)$
        & $0.769_{-0.050}^{+0.013}{}_{-0.039}^{+0.043}{}_{-0.001}^{+0.000}$
        \\
 $\ov B{}^0\to K_S\eta'$
        & $0.60\pm0.07$
        & $(0.691)$
        & $(0.696)$
        & $0.682_{-0.002}^{+0.008}{}_{-0.004}^{+0.004}{}_{-0.000}^{+0.000}$
        \\
        \hline
 $\ov B{}^0\to \pi^+\pi^-$
        & $-0.65\pm 0.07$
        & $(-0.591)$
        & $(-0.533)$
        & $-0.542_{-0.005}^{+0.088}{}_{-0.034}^{+0.038}{}_{-0.074}^{+0.139}$
        \\
 $\ov B{}^0\to \pi^0 \pi^0$
        & --
        & $(0.854)$
        & $(0.820)$
        & $0.484_{-0.114}^{+0.425}{}_{-0.109}^{+0.096}{}_{-0.096}^{+0.145}$
        \\
 $\ov B{}^0\to \eta\eta$
        & --
        & $(-0.985)$
        & $(-0.378)$
        & $-0.308_{-0.237}^{+0.122}{}_{-0.110}^{+0.144}{}_{-0.089}^{+0.160}$
        \\
 $\ov B{}^0\to \eta \eta'$
        & --
        & $(-0.945)$
        & $(-0.956)$
        & $-0.946_{-0.036}^{+0.015}{}_{-0.016}^{+0.020}{}_{-0.016}^{+0.034}$
        \\
 $\ov B{}^0\to \eta'\eta'$
        & --
        & $(-0.901)$
        & $(-0.946)$
        & $-0.917_{-0.024}^{+0.089}{}_{-0.021}^{+0.030}{}_{-0.000}^{+0.001}$
        \\
 $\ov B{}^0\to K^+ K^-$
        & --
        & $(-0.920)$
        & $(-0.468)$
        & $-0.630_{-0.289}^{+0.091}{}_{-0.187}^{+0.521}{}_{-0.046}^{+0.085}$
        \\
 $\ov B{}^0\to K^0\ov K^0$
        & $-0.38^{+0.69}_{-0.77}\pm0.09$
        & $(-0.110)$
        & $(0.184)$
        & $0.327_{-0.283}^{+0.264}{}_{-0.068}^{+0.072}{}_{-0.011}^{+0.002}$
        \\
        & $-1.28^{+0.80}_{-0.73}{}^{+0.11}_{-0.16}$
        \\
 $\ov B{}^0\to \pi^0 \eta$
        & --
        & $(0.019)$
        & (0.064)
        & $0.057_{-0.145}^{+0.151}{}_{-0.012}^{+0.011}{}_{-0.004}^{+0.000}$
        \\
 $\ov B{}^0\to \pi^0\eta'$
        & --
        & $(0.043)$
        & $(-0.011)$
        & $0.084_{-0.124}^{+0.064}{}_{-0.018}^{+0.016}{}_{-0.003}^{+0.0001}$
        \\
         \hline
 $\ov B_s {}^0\to \pi^+\pi^-$
        & --
        & $(0.143)$
        & $(-0.003)$
        & $0.095_{-0.014}^{+0.055}{}_{-0.942}^{+0.109}{}_{-0.001}^{+0.002}$
        \\
 $\ov B_s {}^0\to \pi^0 \pi^0$
        & --
        & $(0.143)$
        & $(-0.003)$
        & $0.095_{-0.014}^{+0.055}{}_{-0.942}^{+0.109}{}_{-0.001}^{+0.002}$
        \\
 $\ov B_s {}^0\to \eta\eta$
        & --
        & $(-0.041)$
        & $(-0.033)$
        & $-0.057_{-0.002}^{+0.029}{}_{-0.017}^{+0.016}{}_{-0.004}^{+0.003}$
        \\
 $\ov B_s {}^0\to \eta \eta'$
        & --
        & $(-0.006)$
        & $(-0.010)$
        & $-0.016_{-0.007}^{+0.016}{}_{-0.003}^{+0.005}{}_{-0.002}^{+0.001}$
        \\
 $\ov B_s {}^0\to \eta'\eta'$
        & --
        & $(0.031)$
        & (0.033)
        & $0.048_{-0.014}^{+0.013}{}_{-0.003}^{+0.003}{}_{-0.000}^{+0.000}$
        \\
 $\ov B_s {}^0\to K^+ K^-$
        & --
        & $(0.194)$
        & (0.202)
        & $0.195_{-0.035}^{+0.019}{}_{-0.021}^{+0.017}{}_{-0.004}^{+0.005}$
        \\
 $\ov B_s {}^0\to K^0\ov K^0$
        & --
        & $(0.005)$
        & (-0.007)
        & $-0.010_{-0.010}^{+0.023}{}_{-0.005}^{+0.007}{}_{-0.002}^{+0.001}$
        \\
 $\ov B_s {}^0\to \pi^0 \eta$
        & --
        & $(0.691)$
        & (0.382)
        & $0.140_{-0.230}^{+0.175}{}_{-0.007}^{+0.008}{}_{-0.025}^{+0.044}$
        \\
 $\ov B_s {}^0\to \pi^0\eta'$
        & --
        & $(0.816)$
        & (0.597)
        & $0.135_{-0.145}^{+0.169}{}_{-0.096}^{+0.095}{}_{-0.037}^{+0.065}$
        \\
        \hline
 $\ov B_s {}^0\to K_S\pi^0$
        & --
        & $(-0.315)$
        & (-0.719)
        & $-0.155_{-0.147}^{+0.116}{}_{-0.047}^{+0.061}{}_{-0.164}^{+0.101}$
        \\
 $\ov B_s {}^0\to K_S\eta$
        & --
        & $(-0.137)$
        & (-0.622)
        & $0.076_{-0.416}^{+0.255}{}_{-0.050}^{+0.031}{}_{-0.157}^{+0.091}$
        \\
 $\ov B_s {}^0\to K_S\eta'$
        & --
        & $(-0.174)$
        & (-0.256)
        & $0.001_{-0.109}^{+0.046}{}_{-0.0848}^{+0.077}{}_{-0.001}^{+0.001}$
        \\
\end{tabular}
\end{ruledtabular}
\end{table}

Time-dependent CP-asymmetries $S$ of most $\ov B{}^0$ decay modes,
except $\ov B{}^0\to \pi^0\pi^0$, $\eta\eta$, $K^+K^-$ and $K^0\ov
K{}^0$ decays, do not receive large contributions from the
residual FSI. Likewise, $S_f$ in most of  $\ov B_s$ modes are
not sensitive to FSI effects, except those in $\ov
B{}^0_s\to\pi^0\eta^{(\prime)}$, $K_S\pi^0$ and
$K_S\eta^{(\prime)}$ decays.

For $\ov B{}^0$ decays, we define $\Delta S\equiv \sin2\beta_{\rm
eff}-\sin2\beta_{c\bar c K}$, where $\sin2\beta_{\rm eff}=-\eta_f
S(f)$ with $\eta_f$ the CP eigenvalue of the state $f$. Comparing
with the recent value of $\sin2\beta_{c\bar c
K}=0.671\pm0.024$~\cite{HFAG} as measured in $B^0\to K$ +
charmonium modes, we obtain:
 \be
 \Delta S(K_S\pi^0)=0.107^{+0.028}_{-0.046},\quad
 \Delta S(K_S\eta)=0.098^{+0.051}_{-0.068},\quad
 \Delta S(K_S\eta')=0.011^{+0.026}_{-0.024}.
 \en
Note that the uncertainty in $\Delta S(K_S\eta')$ are dominated by
the one in the $\sin2\beta_{c\bar cK}$ measurement. The $\Delta
S(K_S\eta')$, being one of the promising tests of the
SM~\cite{Cheng:2005bg}, agrees with the one found in
\cite{Cheng:2005bg,sin2betanew}, while the $\Delta S(K_S\pi^0)$
given here is slightly larger.
The main contribution to the $\chi^2_{\{S(\ov B {}^0)\}}$ given in
Table~\ref{tab:chisquare} is from $S(\ov B {}^0\to K^0 \ov
K{}^0)$.

We note that in the charming penguin approach by Ciuchini et al. ~\cite{RGI}, the
$K\pi$ direct CP violation puzzle $\Delta\A(K\pi)$ can also be
resolved~\cite{CharmingP} and predictions on $\Delta S$ are made.
Ciuchini et al. obtained $\Delta S(K_S\pi^0)=0.024\pm0.059$ and $\Delta
S(K_S\eta')=-0.007\pm0.054$~\cite{CharmingP,review}. Note that (i)
their $\Delta S(K_S\pi^0)$ overlaps with the one given in this
work, (ii) the central value of their $\Delta S(K_S\eta')$ is
negative, but the associated uncertainties allow positive $\Delta
S(K_S\eta')$ as well. There is a considerable overlap between their
$\Delta S(K_S\eta')$ and the one given here.

For $\ov B{}^0_s$ decays, the $S$ contributed from $\ov
B{}^0_s$--$B_s^0$ mixing itself is around $-0.036$. Hence, for
penguin dominated $b\to s$ transition, such as $\ov B_s\to K^0\ov
K{}^0$, $\eta^{(\prime)}\eta^{(\prime)}$ decays, we do not expect
the corresponding $|S|$ to be much larger than ${\cal O}(0.05)$.
Indeed, the predicted $|S|$ as shown in Table~\ref{tab:table-S}
for $\ov B {}^0_s\to \eta\eta$, $\eta'\eta'$ and $\eta\eta'$
decays are all below $0.06$. In particular, given the large $\ov B
{}^0_s\to\eta^{(\prime)}\eta'$ rates, $S(\ov
B_s\to\eta^{(\prime)}\eta')$ are potentially good places to test
the standard model. Given the recent interesting preliminary
results in the $B_s$ phase~\cite{UTfit,PDG}, it will be very
useful to search for $S$ in these $B_s$ charmless decays.

\section{conclusion}

In this work, we study the FSI effects in all charmless $\ov
B_{u,d,s}\to PP$ decay modes. We consider a FSI approach with both
short- and long-distance contributions in which the former are from
all in-elastic channels and are contained in factorization
amplitudes, while the latter are from residual rescattering among
$PP$ states. Flavor SU(3) symmetry is used to constrain the
residual rescattering $S$-matrix. We fit to all available data on
the CP-averaged decay rates and CP asymmetries and make
predictions on yet to be measured ones. Our main results are as follows:

\begin{itemize}

\item Results are in agreement with data in the presence of FSI.

\item The fitted strong phase $\phi_A\simeq-66^\circ$ in
annihilation amplitudes is close to the one used in the S4
scenario of the QCDF approach.

\item For $\ov B$ decays, the $\pi^+\pi^-$ and
$\pi^0\pi^0$ rates are suppressed and enhanced, respectively, by FSI.

\item The deviation ($\Delta\A$) between $\A(\ov B{}^0\to
K^-\pi^+)$ and $\A(B^-\to K^-\pi^0)$ can be understood in the FSI
approach. Since $\A(K^-\pi^0)$ is more sensitive to the residual
rescattering, the degeneracy of these two direct CP violations can
be successfully lifted. However, both short and long distance
strong phases are needed to give correct values for $\A(\ov
K\pi)$s.

\item It is interesting to note that the exchange rescattering is
responsible for generating large and complex color-suppressed
amplitudes, which are crucial in explaining the enhancements in
the $\B(\ov B{}^0\to\pi^0\pi^0)/\B(\ov B{}^0\to\pi^+\pi^-)$ ratio
and the CP asymmetry difference $\Delta\A$.

\item The residual FSI has a large impact on direct CP asymmetries
of many modes.

\item The direct CP violation of $B^-\to \pi^-\pi^0$ is very small
and does not receive any contribution from the residual
rescattering [see Eq.~(\ref{eq:FSIBpipi0})]. It remains as a clean
mode to search for new physics phases.

\item The present data on $\ov B_s\to PP$ decay rates and direct
CP violations can be successfully reproduced.

\item Several $\ov B_s$ decay rates are enhanced by FSI. In particular,
the $\eta'\eta'$ branching ratio  is predicted to reach $10^{-4}$ level, which
can be checked experimentally.

\item Time-dependent CP asymmetry $S$ in $\ov B_{d,s}$ decays
are studied. The $\Delta S(\ov B {}^0\to K_S\eta')$ is very small
($\leq 1\%$). This asymmetry remains as one of the cleanest
measurements to search for new physics phases. The fitted $\Delta
S(\ov B{}^0\to K_S\pi^0)$ is positive and cannot explain the
present $\Delta S(\ov B{}^0\to K_S\pi^0)$ data.

\item Most of the time-dependent CP asymmetries $S$ of $\ov B_s$
to $PP$ states with the strangeness S$=+1$ are expected to be
small. The predicted $|S|$ for $\ov B {}^0_s\to \eta\eta$,
$\eta\eta'$ and $\eta'\eta'$ decays are all below $0.06$. These
modes will be useful to test the SM.

\end{itemize}

\begin{acknowledgments}
The author is grateful to Paoti Chang, Hai-Yang Cheng, Hsiang-nan
Li and Amarjit Soni for helpful discussions. This work is
supported in part by the National Science Council of R.O.C. under
the Grant No. NSC-95-2112-M-033-MY2 and NSC 97-2112-M-033-002-MY3.
\end{acknowledgments}

\appendix

\section{Master Formula of FSI}

Let $H_{\rm W}=\sum_q\lambda_q O_q$ denote the weak decay
Hamiltonian, where $\lambda_q$ are $V_{qb} V^*_{qd}$ (or $V_{qb}
V^*_{qs}$) and $O_q$ are four-quark operators (including Wilson coefficients
$c_i$s). 
From time reversal
invariance of $O_q$, one has,
\begin{eqnarray}
\langle i;\out| O_q|\ov B\rangle^*
 &=&\left(\langle i;\out|\right)^*U^\dagger_T U_T O_{q}^* U^\dagger_T U_T |\ov B\rangle^*
 \nonumber\\
 &=& \langle i;{\rm in}|O_q|\ov B\rangle
 \nonumber\\
 &=& \sum_k\langle i;{\rm in}|\,k;\out\rangle
          \langle k;\out|O_q|\ov B\rangle
 \nonumber\\
 &=& \sum_k\Sc^*_{ki} \langle k;\out|O_q|\ov B\rangle,
 \label{eq:timerev}
\end{eqnarray}
where $\Sc_{ik}\equiv\langle i;\out|\,k;{\rm in}\rangle$ is the
strong interaction $S$-matrix element, and we have used $U_T|{\rm
out\,(in)}\rangle^*=|{\rm in\,(out)}\rangle$ to fix the
phase convention. Eq. (\ref{eq:timerev}) can be solved by (see,
for example \cite{Suzuki:1999uc})
\begin{equation}
 \langle i; \out|O_q|\ov B\rangle=\sum_k\Sc^{1/2}_{ik} A^{q0}_{k},
 \label{eq:FSIsolution}
\end{equation}
where $A^{q0}_{k}$ is a real amplitude. To show that this is
indeed a solution to Eq. (\ref{eq:timerev}), one needs to use
$\Sc_{ik}=\Sc_{ki}$, which follows from the time reversal invariance
of strong interactions and the phase convention we have
adopted. The weak decay amplitude picks up strong scattering
phases \cite{Watson:1952ji} and we have
 \be
 A^{FSI}_i\equiv\langle i; \out|H_{\rm W}|\ov B\rangle
           =\sum_q\langle i;\out|\lambda_q O_q|\ov B\rangle
           =\sum_{q,k}\Sc^{1/2}_{ik} (\lambda_q A^{q0}_{k})
           =\sum_k\Sc^{1/2}_{ik} A^{0}_{k},
 \en
where we have defined $A^0\equiv \sum_q \lambda_q A^{q0}$ free of any strong phase. The above equation is the master
formula for FSI in $\ov B_{u,d,s}$ decays.

\section{Constraints in the U(3) case}

In the U(3) case, one cannot have rescattering from both exchange
and annihilation so that $r^{(m)}_e r^{(m)}_a=0$. This
can be easily seen by inspecting $\Sc^m_{res,3}$ in the
$\pi^-\pi^0-K^0 K^--\pi^- \eta_q-\pi^-\eta_s$ basis, where
$\eta_q=(u\bar u+d\bar d)/\sqrt2$ and $\eta_s=s\bar s$. From
Eq.~(\ref{eq:T123}) and the requirement that $\Sc^m_{res}$ and
$\T^{(m)}$ preserve their forms as determined by U(3) symmetry, we
should have
 \be
 \T^{(m)}_3 &=& \left(
\begin{array}{cccc}
r^{(m)}_0+r^{(m)}_a
      &0
      &0
      &0
      \\
0
      &r^{(m)}_0+r^{(m)}_a
      &\sqrt2 r^{(m)}_a
      &r^{(m)}_e
      \\
0
      &\sqrt2 r^{(m)}_a
      &r^{(m)}_0+2r_a+r^{(m)}_e
      &0
      \\
0
      &r^{(m)}_e
      &0
      &r^{(m)}_0
\end{array}
\right),
 \label{eq:T3U3}
 \en
in the new basis. Under U(3) symmetry, it is evident that
$(\Sc^{m}_{res,3})_{34,43}=0$ for any $m$. Hence, from
 \be
 (\Sc^{2m}_{res,3})_{34,43}=(1+2i \T^{(m)}_3-\T^{(m)}_3\cdot\T_3^{(m)})_{34,43}=-\sqrt2
 r^{(m)}_e r^{(m)}_a,
 \en
which should also be zero, we must have
 \be
 r^{(m)}_e r^{(m)}_a=0
 \en
for any $m$ in the U(3) case.

Given the above constraint, we have two different solutions, which
are,
(a)~annihilation type ($r^{(m)}_a\neq0,\,r^{(m)}_e=0$) and
(b)~exchange type ($r^{(m)}_e\neq0,\,r^{(m)}_a=0$). For the
annihilation type solution, we have
 \be
 1+i(r^{(m)}_0+r^{(m)}_a)
      &=&\frac{2e^{2m\,i\bar\delta}+3e^{2m\,i\delta_8}}{5},
      \nonumber\\
 ir^{(m)}_a
      &=&\frac{3}{5} (e^{2m\,i\delta_8}-e^{2m\,i\bar\delta}),
      \nonumber\\
 i(r^{(m)}_a+r^{(m)}_t)
      &=&\frac{-e^{2m\,i\bar\delta}-4e^{2m\,i\delta_8}+5e^{2m\,i\delta_1}}{20},
      \nonumber\\
      r^{(m)}_e&=&0,
 \label{eq:raU3}
 \en
which corresponds to taking
 \be
 \delta_{27}=\delta_8'=\delta'_1\equiv\bar\delta,
 \quad
 \delta_8,
 \quad
 \delta_1,
 \quad
 \tau=-\frac{1}{2} \sin^{-1}\frac{4\sqrt5}{9},
 \quad
 \nu=-\frac{1}{2}\sin^{-1}\frac{4\sqrt2}{9},
 \label{eq:solutionreU3raB}
 \en
in Eq.~(\ref{eq:solution}).
%
%
For the exchange type solution, we have
 \be
 1+i r^{(m)}_0
      &=&\frac{1}{2} (e^{2m\,i\bar\delta}+e^{2m\,i\delta_8}),
      \nonumber\\
 i r^{(m)}_e
      &=&\frac{1}{2} (e^{2m\,i\bar\delta}-e^{2m\,i\delta_8}),
      \nonumber\\
 r^{(m)}_a
      &=&r^{(m)}_t
      =0,
 \label{eq:reU3}
 \en
which corresponds to setting
 \be
 \delta_{27}=\delta'_8=\delta_1'\equiv\bar\delta,
 \quad
 \delta_8=\delta_1,
 \quad
 \tau=\frac{1}{2} \sin^{-1}\frac{\sqrt5}{3},
 \quad
 \nu=\frac{1}{2}\sin^{-1}\frac{2\sqrt2}{3},
 \label{eq:solutionreU3reB}
 \en
in Eq.~(\ref{eq:solution}).

In the above solutions, we explicitly see that U(3) symmetry
imposes relations on the parameters of different SU(3) multiplets,
and, consequently, reduces the number of independent parameters.
%
%
It should be noted that  mixing angles are
fixed in both solutions.


\begin{thebibliography}{99}

\bibitem{PDG}
C.~Amsler {\it et al.}  [Particle Data Group],
  Phys.\ Lett.\  B {\bf 667}, 1 (2008).
%
%

\bibitem{HFAG}
  E.~Barberio {\it et al.}, [Heavy Flavor Averaging Group (HFAG)],
  arXiv:0808.1297 [hep-ex];
  online update at http://www.slac.stanford.edu/xorg/hfag.



\bibitem{sin2beta}
  Y.~Grossman and M.~P.~Worah,
  Phys.\ Lett.\  B {\bf 395}, 241 (1997)
  [arXiv:hep-ph/9612269];
  D.~London and A.~Soni,
  Phys.\ Lett.\  B {\bf 407}, 61 (1997)
  [arXiv:hep-ph/9704277];
  Y.~Grossman, G.~Isidori and M.~P.~Worah,
  Phys.\ Rev.\  D {\bf 58}, 057504 (1998)
  [arXiv:hep-ph/9708305];
  Y.~Grossman, Z.~Ligeti, Y.~Nir and H.~Quinn,
  Phys.\ Rev.\  D {\bf 68}, 015004 (2003)
  [arXiv:hep-ph/0303171];
  M.~Gronau, Y.~Grossman and J.~L.~Rosner,
  Phys.\ Lett.\  B {\bf 579}, 331 (2004)
  [arXiv:hep-ph/0310020];
  M.~Gronau, J.~L.~Rosner and J.~Zupan,
  Phys.\ Lett.\  B {\bf 596}, 107 (2004)
  [arXiv:hep-ph/0403287].

\bibitem{Cheng:2005bg}
  H.~Y.~Cheng, C.~K.~Chua and A.~Soni,
  Phys.\ Rev.\  D {\bf 72}, 014006 (2005)
  [arXiv:hep-ph/0502235].

\bibitem{sin2betanew}
  M.~Beneke,
  Phys.\ Lett.\  B {\bf 620}, 143 (2005)
  [arXiv:hep-ph/0505075].

\bibitem{Chua:2006hr}
  See aslo, C.~K.~Chua,
{\it In the Proceedings of 4th Flavor Physics and CP Violation
Conference (FPCP 2006), Vancouver, British Columbia, Canada, 9-12
Apr 2006, pp 008}
  [arXiv:hep-ph/0605301];
{\it In the Proceedings of 6th Flavor Physics and CP Violation
Conference (FPCP 2008), Taipei, Taiwan, 5-9 May 2008}
  arXiv:0807.3596 [hep-ph].



\bibitem{Gronau:2006qh}
  M.~Gronau, J.~L.~Rosner and J.~Zupan,
  Phys.\ Rev.\  D {\bf 74}, 093003 (2006)
  [arXiv:hep-ph/0608085].

\bibitem{review}
M.~Artuso {\it et al.},
  arXiv:0801.1833 [hep-ph].

\bibitem{pQCD}
  Y.~Y.~Keum, H.~n.~Li and A.~I.~Sanda,
  Phys.\ Lett.\  B {\bf 504}, 6 (2001)
  [arXiv:hep-ph/0004004];
%
  Phys.\ Rev.\  D {\bf 63}, 054008 (2001)
  [arXiv:hep-ph/0004173].

\bibitem{Beneke:2001ev}
  M.~Beneke, G.~Buchalla, M.~Neubert and C.~T.~Sachrajda,
  Nucl.\ Phys.\  B {\bf 606}, 245 (2001)
  [arXiv:hep-ph/0104110].

\bibitem{Buras:2004ub}
  A.~J.~Buras, R.~Fleischer, S.~Recksiegel and F.~Schwab,
  Nucl.\ Phys.\  B {\bf 697}, 133 (2004)
  [arXiv:hep-ph/0402112].

\bibitem{Buras:2004th}
  A.~J.~Buras, R.~Fleischer, S.~Recksiegel and F.~Schwab,
  Acta Phys.\ Polon.\  B {\bf 36}, 2015 (2005)
  [arXiv:hep-ph/0410407].

\bibitem{Buras:2005cv}
  A.~J.~Buras, R.~Fleischer, S.~Recksiegel and F.~Schwab,
  Eur.\ Phys.\ J.\  C {\bf 45}, 701 (2006)
  [arXiv:hep-ph/0512032].



\bibitem{Li:2005kt}
  H.~n.~Li, S.~Mishima and A.~I.~Sanda,
  Phys.\ Rev.\  D {\bf 72}, 114005 (2005)
  [arXiv:hep-ph/0508041].

\bibitem{CharmingP}
M. Pierini, talk given at The 2007 Europhysics Conference on High
Energy Physics (EPS2007),  19-25 July  2007, Manchester, England;
M. Ciuchini et al., in preparation.

\bibitem{Baek2005}
S.~Baek, P.~Hamel, D.~London, A.~Datta and D.~A.~Suprun,
Phys.\ Rev.\ D {\bf 71}, 057502 (2005) [arXiv:hep-ph/0412086];
S.~Baek,
JHEP {\bf 0607}, 025 (2006) [arXiv:hep-ph/0605094];
S.~Baek and D.~London,
Phys.\ Lett.\ B {\bf 653}, 249 (2007) [arXiv:hep-ph/0701181].



\bibitem{Hou:2006jy}
  W.~S.~Hou, H.~n.~Li, S.~Mishima and M.~Nagashima,
  Phys.\ Rev.\ Lett.\  {\bf 98}, 131801 (2007)
  [arXiv:hep-ph/0611107].

\bibitem{Kim:2007kx}
  C.~S.~Kim, S.~Oh and Y.~W.~Yoon,
  arXiv:0707.2967 [hep-ph].

\bibitem{Nature}
S.-W. Lin {\it et al.}, Belle Collaboration, Nature 452, 332
(2008).

\bibitem{Hou:1999st}
  W.~S.~Hou and K.~C.~Yang,
  Phys.\ Rev.\ Lett.\  {\bf 84}, 4806 (2000)
  [Erratum-ibid.\  {\bf 90}, 039901 (2003)]
  [arXiv:hep-ph/9911528].

\bibitem{Chua:2001br}
C.K.~Chua, W.S.~Hou and K.C.~Yang,
Phys.\ Rev.\ D {\bf 65}, 096007 (2002) [hep-ph/0112148];
  C.~K.~Chua and W.~S.~Hou,
  Phys.\ Rev.\  D {\bf 72}, 036002 (2005)
  [arXiv:hep-ph/0504084];
  C.~K.~Chua and W.~S.~Hou,
  Phys.\ Rev.\  D {\bf 77}, 116001 (2008)
  [arXiv:0712.1882 [hep-ph]].




\bibitem{Cheng:2004ru}
  H.Y.~Cheng, C.K.~Chua and A.~Soni,
  Phys.\ Rev.\ D {\bf 71}, 014030 (2005)
  [hep-ph/0409317].


\bibitem{UTfit}
  M.~Bona {\it et al.}  [UTfit Collaboration],
  arXiv:0803.0659 [hep-ph].
  Diego Tonelli, talk given at ICHEP2008, July 29th. - August 5th.,
  2008, Philadelphia;
  John Ellison, talk given at ICHEP2008, July 29th. - August 5th.,
  2008, Philadelphia.


\bibitem{Beneke:2003zv}
  M.~Beneke and M.~Neubert,
  Nucl.\ Phys.\  B {\bf 675}, 333 (2003)
  [arXiv:hep-ph/0308039];
M.~Beneke and M.~Neubert,
  Nucl.\ Phys.\  B {\bf 651}, 225 (2003)
  [arXiv:hep-ph/0210085].



\bibitem{SCET}
  C.~W.~Bauer, S.~Fleming, D.~Pirjol and I.~W.~Stewart,
  Phys.\ Rev.\  D {\bf 63}, 114020 (2001)
  [arXiv:hep-ph/0011336];
  C.~W.~Bauer, D.~Pirjol, I.~Z.~Rothstein and I.~W.~Stewart,
  Phys.\ Rev.\  D {\bf 70}, 054015 (2004)
  [arXiv:hep-ph/0401188].


\bibitem{quasielastic0}
W.S. Hou, private communication.

\bibitem{quasielastic}
C.~K.~Chua, W.~S.~Hou and K.~C.~Yang,
  Mod.\ Phys.\ Lett.\  A {\bf 18}, 1763 (2003)
  [arXiv:hep-ph/0210002].




\bibitem{Smith}
  C.~Smith,
  Eur.\ Phys.\ J.\  C {\bf 10}, 639 (1999)
  [arXiv:hep-ph/9808376];
  Eur.\ Phys.\ J.\  C {\bf 33}, 523 (2004)
  [arXiv:hep-ph/0309062].



\bibitem{Donoghue:1996hz}
  J.~F.~Donoghue, E.~Golowich, A.~A.~Petrov and J.~M.~Soares,
  Phys.\ Rev.\ Lett.\  {\bf 77}, 2178 (1996)
  [arXiv:hep-ph/9604283].

\bibitem{Zenczykowski:2003ed}
  P.~Zenczykowski and P.~Lach,
  Phys.\ Rev.\  D {\bf 69}, 094021 (2004)
  [arXiv:hep-ph/0309198].




\bibitem{Suzuki:2007je}
  M.~Suzuki,
  arXiv:0710.5534 [hep-ph].


\bibitem{Wu:2007tm}
  Y.~L.~Wu, Y.~F.~Zhou and C.~Zhuang,
  arXiv:0712.2889 [hep-ph].


\bibitem{RGI}
M.~Ciuchini, E.~Franco, G.~Martinelli and L.~Silvestrini,
  Nucl.\ Phys.\  B {\bf 501}, 271 (1997)
  [arXiv:hep-ph/9703353];
A.~J.~Buras and L.~Silvestrini,
  Nucl.\ Phys.\  B {\bf 569}, 3 (2000)
  [arXiv:hep-ph/9812392].


\bibitem{Feldmann:1998vh}
T.~Feldmann, P.~Kroll and B.~Stech,
Phys.~Rev.~D {\bf 58}, 114006 (1998) [hep-ph/9802409];
Phys.\ Lett.\ {\bf B449}, 339 (1999) [hep-ph/9812269].

\bibitem{TDLee}
  T.~D.~Lee,
  Contemp.\ Concepts Phys.\  {\bf 1}, 1 (1981).

\bibitem{Pham:2007nt}
 T.~N.~Pham,
  Phys.\ Rev.\  D {\bf 77}, 014024 (2008)
  [Erratum-ibid.\  D {\bf 77}, 019905 (2008)]
  [arXiv:0710.2412 [hep-ph]].



\bibitem{Petadata}
 B.~Aubert {\it et al.}  [BABAR Collaboration],
  arXiv:0804.2422 [hep-ex];
 W. Ford, talk given at FPCP2008, May 5-9th., 2008, Taipei;
  B.~Aubert {\it et al.}  [BABAR Collaboration],
  arXiv:0807.4226 [hep-ex].



\bibitem{CDFnew}
A. Warburton, talk given at 6th Flavor Physics and CP Violation
Conference (FPCP2008), 5-9 May 2008, Taipei.


\bibitem{CKMfitter}
  J.~Charles {\it et al.}  [CKMfitter Group],
  Eur.\ Phys.\ J.\  C {\bf 41}, 1 (2005)
  [arXiv:hep-ph/0406184];
  online update at http://ckmfitter.in2p3.fr/.


\bibitem{Gegenbauer}
P.~Ball, V.~M.~Braun and A.~Lenz,
  JHEP {\bf 0605}, 004 (2006)
  [arXiv:hep-ph/0603063];
G.~Duplancic, A.~Khodjamirian, T.~Mannel, B.~Melic and N.~Offen,
  JHEP {\bf 0804}, 014 (2008)
  [arXiv:0801.1796 [hep-ph]].


\bibitem{LF}
H.Y.~Cheng, C.K.~Chua and C.W.~Hwang,
  Phys.\ Rev.\ D {\bf 69}, 074025 (2004)
  [hep-ph/0310359].

\bibitem{MS}
  D.~Melikhov and B.~Stech,
  Phys.\ Rev.\ D {\bf 62}, 014006 (2000)
  [arXiv:hep-ph/0001113].



\bibitem{largeC}
C.~W.~Chiang, M.~Gronau, J.~L.~Rosner and D.~A.~Suprun,
  Phys.\ Rev.\  D {\bf 70}, 034020 (2004)
  [arXiv:hep-ph/0404073];
Y.~Y.~Charng and H.~n.~Li,
  Phys.\ Rev.\  D {\bf 71}, 014036 (2005)
  [arXiv:hep-ph/0410005].



\bibitem{Ishino:2006if}
  H.~Ishino {\it et al.}  [Belle Collaboration],
  Phys.\ Rev.\ Lett.\  {\bf 98}, 211801 (2007)
  [arXiv:hep-ex/0608035].


\bibitem{CPT}
 J.~M.~Gerard and W.~S.~Hou,
  Phys.\ Rev.\ Lett.\  {\bf 62}, 855 (1989);
D.~Atwood, S.~Bar-Shalom, G.~Eilam and A.~Soni,
  Phys.\ Rept.\  {\bf 347}, 1 (2001)
  [arXiv:hep-ph/0006032].



\bibitem{Aubert:2006gm}
  B.~Aubert {\it et al.}  [BABAR Collaboration],
  Phys.\ Rev.\ Lett.\  {\bf 97}, 171805 (2006)
  [arXiv:hep-ex/0608036].

\bibitem{Abe:2007xd}
  K.~Abe {\it et al.}  [Belle Collaboration],
  arXiv:0708.1845 [hep-ex].



\bibitem{Suzuki:1999uc}
M.~Suzuki and L.~Wolfenstein,
Phys.\ Rev.\ D {\bf 60}, 074019 (1999) [hep-ph/9903477].

\bibitem{Watson:1952ji}
K.~M.~Watson,
Phys.\ Rev.\  {\bf 88}, 1163 (1952).










\end{thebibliography}
\end{document}